\documentclass[preprint,review,12pt]{elsarticle}

\usepackage{graphicx}
\usepackage{amssymb}
\usepackage{amsmath,bm}
\usepackage{colortbl,booktabs}
\usepackage{tabularx}
\usepackage{threeparttable}
\usepackage{multicol,multirow}
\usepackage{subfigure}
\usepackage [font=footnotesize,labelfont=bf]{caption}
\usepackage{rotating}
\usepackage{hyperref}
\usepackage{lscape}
\makeatletter
\renewcommand{\@thesubfigure}{\hskip\subfiglabelskip}
\makeatother
\usepackage{color}
\definecolor{R}{rgb}{1, 0, 0}
\definecolor{G}{rgb}{0, 0.5, 0}
\definecolor{B}{rgb}{0, 0, 1}

\definecolor{yjzhuCol}{rgb}{0.9, 0, 0.2}
\definecolor{czhangCol}{rgb}{0.9, 0, 0.2}
\definecolor{xyhuCol}{rgb}{0.9, 0, 0.9}

\journal{Elsevier}

\begin{document}

\begin{frontmatter}

\title{An efficient and generalized consistency correction method for weakly-compressible SPH}
%\tnotetext[label0]{This is only an example}

\author[mymainaddress,mysecondaryaddress]{Yaru Ren}
%\ead{2019323060052@scu.stu.edu.cn}

\author[mymainaddress]{Pengzhi Lin} 
%\ead{cvelinpz@scu.edu.cn}

\author[mysecondaryaddress,mythirdaddress]{Chi Zhang}
%\ead{c.zhang@tum.de}

\author[mysecondaryaddress]{Xiangyu Hu \corref{mycorrespondingauthor}}
\cortext[mycorrespondingauthor]{Corresponding author.}
\ead{xiangyu.hu@tum.de}

\address[mymainaddress]{State Key Laboratory of Hydraulics and Mountain River Engineering, Sichuan University, Chengdu, Sichuan, China}
\address[mysecondaryaddress]{Department of Mechanical Engineering, Technical University of Munich\\
85748 Garching, Germany}
\address[mythirdaddress]{Huawei Technologies Munich Research Center\\
80992 Munich, Germany}

\begin{abstract}
In this paper, 
a new efficient and generalized consistency correction method for weakly-compressible 
smoothed particle hydrodynamics is proposed and successfully implemented in the simulation of violent free-surface flow 
exhibiting breaking and impact events for the first time. 
It's well known that the original kernel gradient correction (KGC) encounters numerical instability resulting from matrix inversion.  
The present method remedies this issue by introducing a weighted average of the KGC matrix and the identity matrix,
other than directly applying  KGC matrix, 
to achieve numerical stability meanwhile decrease numerical dissipation. 
To ensure momentum conservation, 
the correction is implemented in a particle-average pattern by rewriting 
the the pressure term of the Riemann solution. 
Furthermore, 
the proposed weighted KGC scheme  is incorporated into the dual-criteria time-stepping framework  developed by Zhang et al. (2020) \cite{22}
to achieve optimized computational efficiency. 
A set of numerical examples in both two- and three-dimensions are investigated to demonstrate that the present method can significantly reduce numerical dissipation meanwhile exhibit a smooth pressure field for general free-surface flows. 
\end{abstract}


\end{frontmatter}
% \linenumbers
%%%%%%%%%%%%%%%%%%%%%%%%%%%%%%%%%%%%%%%%%%%%%%%%%%%%%%%%%%%%%
%
% Section
%
%%%%%%%%%%%%%%%%%%%%%%%%%%%%%%%%%%%%%%%%%%%%%%%%%%%%%%%%%%%%%
\section{Introduction}
\label{Introduction}
In the past decades, particle-based methods have attracted more and more attentions thanks to their Lagrangian nature which can easily handle large material deformations and capture moving surfaces and interfaces. As one of the key examples, smoothed particle hydrodynamics (SPH) \cite{1,2,3,67} has experienced tremendous developments in engineering applications, such as those of coastal and ocean engineering \cite{2,68}, astrophysics \cite{57}, geotechnical engineering \cite{56}, and so on. 
While recognized as a promising potential in coastal and ocean engineering, 
the SPH method encounters excessive numerical dissipation in the simulation of the wave propagation \cite{4,27}, 
leading to the over damping of mechanical energy \cite{6,5}. 

To remedy this issue, several algorithms have been developed in the literature and they are generally categorized into three groups, 
i.e. introducing adaptive numerical dissipation, increasing the smoothing length and applying kernel gradient correction (KGC). 
The adaptive numerical dissipation scheme, including $\delta$-SPH \cite{50,28} and Riemann-SPH \cite{21,30}, has demonstrated its  ability to improve the energy conservation 
while still exhibit over damping of mechanical energy in wave dynamics. 
Increasing the smoothing length is a simple and effective approach for reducing the numerical dissipation in the weekly-compressible SPH (WCSPH) method. 
Typically, the smoothing length is set to be twice the initial particle spacing to achieve satisfactory results, as reported in the literature\cite{7,28,30}. However, this approach also incurs a significantly higher computational cost, particularly in 3D simulations.
%%%%%%%%%%%%%%%%%%%%%%%%%%%%%%%%%%%%%%%%%%%%%%%%%
\begin{table}[t]
  \centering
  \caption{Different KGC schemes in literature.}
  \begin{threeparttable} 
  \resizebox{\textwidth}{!}{
  \begin{tabularx}{1.2\textwidth}{p{4em}<{\centering}p{18em}<{\centering}p{6em}<{\centering}p{6em}<{\centering}}
    \hline
    No. & Corrected gradient form \tnote{1}  & Momentum Conservation & Numerical Stability  \\
    \hline 
     S1 &$\widetilde{\nabla }_iW_{ij}=\mathbf{B}_{i}\nabla_iW_{ij}$ & No & No\\
     S2 &$\widetilde{\nabla }_iW_{ij}=\frac{1}{2}\left(\mathbf{B}_{i}+\mathbf{B}_{j}\right)\nabla_iW_{ij}$ & Yes & No\\
     S3 &$\widetilde{\nabla }_iW_{ij}=\frac{1}{2}\left(\mathbf{A}_{i}+\mathbf{A}_{j}\right)^{-1}\nabla_iW_{ij}$ & Yes & No\\
     S4 &$\widetilde{\nabla }_iW_{ij}= diag\left(\mathbf{B}_{i}\right)\nabla_iW_{ij}$ & No & No\\
   \hline
\end{tabularx}        
}
\begin{tablenotes}   
\footnotesize               
\item[1] $\mathbf{B}$ the correction matrix ($\mathbf{B} = \mathbf{A}^{-1}$), $\nabla W$ the original kernel gradient and $\widetilde{\nabla } W$ the \\
corrected kernel gradient. Refer to Sections \ref{Weakly compressible SPH method} and \ref{Weighted kernel gradient correction} for more details. 
\end{tablenotes}  
 \end{threeparttable} 
\label{table: KGC}
\end{table}
%%%%%%%%%%%%%%%%%%%%%%%%%%%%%%%%%%%%%%%%%%%%%%%%%

The KGC scheme, 
also known as renormalized scheme, 
was first investigated by Randles et al. \cite{13}. 
Since then, it has been extensively studied and incorporated into different SPH methods to improve numerical accuracy and consistency \cite{14,15,4,16}. 
Several formulations have been proposed in the literature concerning the proper implementation of the KGC, 
which are briefly summarized in Table \ref{table: KGC}. 
Specifically, 
scheme  S1, 
encounters two drawbacks \cite{17,16}, 
viz, the momentum non-conservation duo to its asymmetric form and the numerical instability as the ill-conditioned matrix inversion of particles close to boundary. 
To address these issues, 
Vila \cite{66} suggested a symmetric version as scheme S2 which preserves momentum conservation while  is not first-order consistent any more. 
Guilcher et al.\cite{4} applied this scheme to wave dynamics and achieved improved performance in predicting wave propagation. 
Zago et al. \cite{5} pointed out that this scheme exhibit numerical instability in the proximity of the free surface when a smaller artificial viscosity coefficient is applied. 
To address this instability, 
Zago et al. \cite{5} proposed a new symmetric version by first obtaining the particle average matrix and then its inverse as scheme S3.  
Even though scheme S3 exhibits improved robust features, it requires calculating the inverse matrix for every particle pair, 
introducing extra computational efforts \cite{18} compared with scheme S2. 
Zhang and Liu \cite{20} introduced scheme S4 by assuming 
that the correction matrix is diagonally dominant and therefore the contributions from other directions can be neglected to avoid the correction matrix inversion. 
However, this assumption is not reasonable for violent free-surface flow 
exhibiting breaking and impact events, 
leading to the loss of accuracy. 
Instead of applying a KGC scheme to all fluid particles, 
Zago et al. \cite{5} achieved stable simulation by not applying the correction to particles with incomplete supports or distorted configurations. 
They introduced a cut-off threshold based on the determinant value of the correction matrix to detect particles with matrix deficiencies. 
One main weakness is that the tunable threshold requires careful numerical calibrations and its values usually are case dependent, 
in particular for modeling violent free-surface flows. 
Therefore, introducing KGC to increase accuracy and decrease numerical dissipation without inducing numerical instability for general free surface flows is still not addressed in the literature.

In this paper, we focus on the KGC scheme and propose a generalized and consistent weighted KGC (WKGC)  for WCSPH to decrease the numerical dissipation meanwhile guarantee numerical stability for general free surface flows. Instead of directly applying the KGC or introducing a cut-off threshold, the present method introduces a weighted average of the original KGC matrix and the identity matrix to address the induced numerical instability. Therefore, the particles with ill-conditioned correction matrices are corrected by diagonally dominant matrices to ensure numerical stability. 
Then, 
the correction is implemented in a particle-average pattern by rewriting 
the the pressure term of the Riemann solution to ensure momentum conservation. 
Also, the WKGC is incorporated into a dual-criteria time stepping framework, taking into account computational efficiency. Two- and three-dimensional cases, including standing wave, oscillation drop, dam break and 3D wave-structure interaction, are investigated to test the accuracy and stability of the method. 
%%%%%%%%%%%%%%%%%%%%%%%%%%%%%%%%%%%%%%%%%%%%%%%%%%%%%%%%%%%%%
%
% Section
%
%%%%%%%%%%%%%%%%%%%%%%%%%%%%%%%%%%%%%%%%%%%%%%%%%%%%%%%%%%%%%
\section{Weakly compressible SPH method}
\label{Weakly compressible SPH method}
\subsection{Governing equation}
The governing equation in the lagrangian framework for viscous flows includes the mass and momentum conservation equations read:
\begin{equation}
\left\{\begin{aligned}
\frac{d\rho }{dt}&=-\rho\nabla \cdot \textbf{v} \\
\frac{d\textbf{v} }{dt}&=-\frac{1}{\rho}\nabla p + \textbf{a}_\nu + \textbf{g} ,\label{eq1}\\
\end{aligned}\right.
\end{equation}
where $\rho$ the fluid density, $\textbf{v}$ the velocity, $p$ the pressure, $\textbf{g}$ the gravitational acceleration, $\textbf{a}_\nu$ the acceleration due to the  viscous force and $\frac{d}{dt} = \frac{\partial}{\partial t} + \textbf{v}\cdot\nabla$ refers to the material derivative.

To model the incompressible flow with the weakly-compressible assumption, the pressure and density are related through an artificial equation of state (Eos)
\begin{equation}
 p=c_0^2(\rho - \rho_0) , \label{eq2}\\
\end{equation}
where $\rho_0$ the initial density and $c_0$ the artificial speed of sound. With the weakly-compressible assumption, the density variation maintains below 1\% by setting $c_0=10U_{max}$ with $U_{max}$ denoting the anticipated maximum fluid speed.
%%%%%%%%%%%%%%%%%%%%%%%%%%%%%%%%%%%%%%%%%%%%%%%%%%%%%%%%%%%%%
% Section
%%%%%%%%%%%%%%%%%%%%%%%%%%%%%%%%%%%%%%%%%%%%%%%%%%%%%%%%%%%%%
\subsection{Riemann-SPH method}

To discretize the system of Eq.(\ref{eq1}), we adopt the Riemann-SPH method \cite{21} where the continuity and momentum equations are discretized as
\begin{equation}
\left\{\begin{aligned}
\frac{d\rho_i}{dt}&=2{\rho_i}\sum \limits_{j}(\textbf{v}-\textbf{v}^{*})\cdot V_j\nabla_iW_{ij}\\
\frac{d\textbf{v}_i}{dt}&=-2\sum \limits_{j}\frac{{P}^{*}}{\rho_i}V_j\nabla_iW_{ij} + 2\sum \limits_{j}\frac{\nu}{\rho_i}\textbf{v}_{ij}V_j \nabla_iW_{ij} + \textbf{g}_i  , \label{eq3}\\
\end{aligned}\right.
\end{equation}
where $V_j$ is the particle volume, $\textbf{v}_{ij}=\textbf{v}_i-\textbf{v}_j$ the relative velocity. $\nu$ the fluid kinetic viscosity and $\nabla_iW_{ij} = \frac{\partial W}{\partial \textbf{r}_{ij}}\textbf{e}_{ij}$ with $\textbf{e}_{ij}= \frac{\textbf{r}_{ij}}{\left |\textbf{r}_{ij}\right |}$. Note that, $\textbf{v}^*$ and $P^*$ are the solutions of the Riemann problem constructed along the interacting line of a particle pair pointed from particle $i$ to $j$. The left and the right states of the Riemann problem are defined
\begin{equation}
\left\{\begin{aligned}
\left(\rho_L,U_L,P_L,c_L\right)&= \left(\rho_i,-\textbf{v}_i\cdot \textbf{e}_{ij},p_i,c_i\right)\\ 
\left(\rho_R,U_R,P_R,c_R\right)&= \left(\rho_j,-\textbf{v}_j\cdot \textbf{e}_{ij},p_j,c_j\right) , \label{eq4}
\end{aligned}\right.
\end{equation}
To solve the Riemann problem, we apply the linearised Riemann solver with a dissipation limiter \cite{21} and obtain
\begin{equation}
\left\{\begin{aligned}
\textbf{v}^{*}&=\frac{1}{2}(\textbf{v}_i+\textbf{v}_j)-(U^{*}-\frac{1}{2}(U_L+U_R))\cdot\textbf{e}_{ij}\\
U^{*}&=\frac{(\rho_Lc_LU_L+\rho_Rc_RU_R+P_L-P_R)}{\rho_Lc_L+\rho_Rc_R}\\
P^{*}&=\frac{\left(\rho_Lc_LP_R+\rho_Rc_RP_L+\rho_Lc_L\rho_Rc_R\left(U_L-U_R\right)\beta\right)}{\rho_Lc_L+\rho_Rc_R} ,\label{eq5}\\
\end{aligned}\right.
\end{equation}
where $\beta=min\left(3max\left((U_L-U_R )/\left((c_L-c_R)/(\rho_L+\rho_R)\right),0.0\right),1.0\right)$ is the low-dissipation limiter. 

%%%%%%%%%%%%%%%%%%%%%%%%%%%%%%%%%%%%%%%%%%%%%%%%%%%%%%%%%%%%%
% Section
%%%%%%%%%%%%%%%%%%%%%%%%%%%%%%%%%%%%%%%%%%%%%%%%%%%%%%%%%%%%%
\section{Weighted kernel gradient correction}
\label{Weighted kernel gradient correction}
In this part, the WKGC scheme is presented in detail with its implementation in Riemann-SPH and dual-criteria time stepping frameworks.

%%%%%%%%%%%%%%%%%%%%%%%%%%%%%%%%%%%%%%%%%%%%%%%%%%%%%%%%%%%%%%%%%%%%%%%%
\subsection{Weighted KGC scheme} 

In SPH, the kernel approximation of the gradient of a field reads 
\begin{equation}
\nabla f(\mathbf{r})=\int_{\Omega}\left(f(\mathbf{r'})-f(\mathbf{r})\right)\nabla W d\mathbf{r'} .\label{eq9}\\
\end{equation}

Then we can conduct the Taylor expansion and have
\begin{equation}
\nabla f(\mathbf{r})=\nabla f(\mathbf{r})\int_{\Omega}\left(\mathbf{r'}-\mathbf{r}\right)\otimes\nabla W d\mathbf{r'}+ O(h^2) .\label{eq10}\\
\end{equation}

It is easy to conclude that Eq.(\ref{eq10}) achieves 1-order consistency if the following condition is satisfied
\begin{equation}
\int_{\Omega}\left(\mathbf{r'}-\mathbf{r}\right)\otimes\nabla W d\mathbf{r'}=\mathbf{I} .\label{eq11}\\
\end{equation}

In term of particle approximation \cite{43}, Eq.(\ref{eq11}) can be rewritten as
\begin{equation}
\sum \limits_{j}\mathbf{r}_{ji}\otimes \nabla_iW_{ij}V_j=\mathbf{I} .\label{eq12}\\
\end{equation}
%Where $\mathbf{r}_{ji}=\mathbf{r}_{j}-\mathbf{r}_{i}$.

However, Eq.(\ref{eq12}) can't be fulfilled for the irregular particle distribution or particles close to the boundary. Therefore, in the KGC scheme, a corrected matrix $\mathbf{B}_i$  is introduced to modify Eq.(\ref{eq12}) as
\begin{equation}
\mathbf{B}_i\sum \limits_{j}\mathbf{r}_{ji}\otimes \nabla_iW_{ij}V_j=\mathbf{I} ,\label{eq13}\\
\end{equation}
where
\begin{equation}
\mathbf{B}_i=\left(\sum \limits_{j}\mathbf{r}_{ji}\otimes \nabla_iW_{ij}V_j\right)^{-1} = \left(\mathbf{A}_i\right)^{-1}\label{eq14}\\
\end{equation}
The KGC scheme can improve computational accuracy and achieve 1st-order consistency both for irregular particle distributions and particles close to the boundary. While it may violate the anti-symmetric property of the SPH discretization, indicating a non-conservation form of momentum. Moreover, the KGC suffers instability when the condition number of the matrix $\mathbf{A}$ is very large \cite{24}. More precisely, when the matrix determinant tends to be zero, a small disturbance may result in large changes in the inverse matrix and induces numerical instability. As discussed in Section \ref{Introduction}, the existing variants of the KGC still suffer from the numerical instability issue.

Inspired by the fact that the identity matrix correction, viz, without any correction, can be introduced to particles with ill-conditioned correction matrices, we introduce a non-linear weight between the original KGC and the identity one to automatically adjust the amount of correction by considering the matrix determinant value. To this end, the weighted KGC is rewritten as
\begin{equation}
\widetilde{\mathbf{B}_i}=\frac{\left | \mathbf{B}_i\right |^\beta\mathbf{B}_i}{\alpha\left |\mathbf{I}\right |+\left |\mathbf{B}_i\right |^\beta} + \frac{\alpha\mathbf{I}}{\alpha\left |\mathbf{I}\right |+\left |\mathbf{B}_i\right |^\beta} ,\label{eq18}\\
\end{equation}
here $\left| ~\right| $ denotes the determinant value, $\alpha$ and $\beta$ are weighting factors, where $\alpha$ is a positive value and $\beta$ is an integer. The weighted matrix becomes the original KGC if $\alpha = 0$. While the first term to enhance accuracy is dominant when the matrix determinant $\left |\mathbf{B}\right |$ is large, the identity matrix is dominant when $\left |\mathbf{B}\right |$ is small to avoid numerical instability. The preliminary numerical tests indicate that the values of $\alpha = 0.3$ and $\beta = 2$ are generally effective parameters.

%%%%%%%%%%%%%%%%%%%%%%%%%%%%%%%%%%%%%%%%%%%%%%%%%%%%%%%%%%%%%
% Section
%%%%%%%%%%%%%%%%%%%%%%%%%%%%%%%%%%%%%%%%%%%%%%%%%%%%%%%%%%%%%
\subsection{Corrected Riemann solution}

To reference the previous study \cite{19}, we only introduce the WKGC scheme into the particle-average term of the momentum equation. Therefore, it is necessary to further decompose the Riemann solution of the pressure term in Eq.(\ref{eq3}) into the inter-particle average and dissipation terms as
\begin{equation}
\left\{\begin{aligned}
\frac{d\textbf{v}_i}{dt}&=-2\sum \limits_{j}\left(\frac{{P}^{*}}{\rho_i}\right)V_j\nabla_iW_{ij}+ 2\sum\limits_{j}\frac{\nu}{\rho_i}\textbf{v}_{ij}V_j \nabla_iW_{ij} + \textbf{g}_i \\
{P}^{*}&= {p}^{*}+{\Pi}^{*}  ,\label{eq6}\\
\end{aligned}\right.
\end{equation}
where $p^*= \frac{\rho_Lc_LP_R+\rho_Rc_RP_L}{\rho_Lc_L+\rho_Rc_R}$ and $\Pi^*= \frac{\rho_Lc_L\rho_Rc_R\left(U_L-U_R\right)\beta}{\rho_Lc_L+\rho_Rc_R}$. To ensure momentum conservation, we implement the WKGC scheme in the particle-average pattern of Eq.(\ref{eq6}) as
\begin{equation}
p^*= \frac{\rho_{L}c_{L}P_{R} \mathbf{B}_{i}+\rho_{R}c_{R}P_{L}\mathbf{B}_{j}}{\rho_{L}c_{L}+\rho_{R}c_{R}}. \label{eq19}
\end{equation}

%%%%%%%%%%%%%%%%%%%%%%%%%%%%%%%%%%%%%%%%%%%%%%%%%%%%%%%%%%%%%
% Section
%%%%%%%%%%%%%%%%%%%%%%%%%%%%%%%%%%%%%%%%%%%%%%%%%%%%%%%%%%%%%
\subsection{Dual-criteria time stepping}
As the main drawback of introducing the KGC scheme is the induced computational efforts, we adopt the dual-criterial time-stepping to increase the computational efficiency. The dual-criterial time stepping proposed by Zhang et al.\cite{22} consists of an advection time step and an acoustic one. The advection criterion is based on the flow velocity and controls the update frequency of the particle neighbor lists or configurations. On the other hand, the acoustic criterion is determined by the artificial speed of sound and is used for the time integration of the governing equations. Generally, the advection time step is larger than the acoustic time step, implying that several acoustic time steps are carried out in one advection time step without updating the particle configuration.
Following Ref.\cite{22}, the advection and acoustic criteria are given by 
\begin{equation}
\left\{\begin{aligned}
\Delta t_{ad}&=0.25\frac{h}{\left|\mathbf{v}\right|_{max}}\\
\Delta t_{ac}&=0.6\left(\frac{h}{c+\left|\mathbf{v}\right|_{max}}\right) , \label{eq20}
\end{aligned}\right.
\end{equation}
where $h$ is the smoothing length,$\left|\mathbf{v}\right|_{max}$ is the maximum particle velocity. 

In this study, we integrate the WKGC scheme into the dual-criterial time stepping framework. 
The key idea is to compute the WKGC matrix and conduct the corresponding inversion for all particles at the advection step and then we use it for several acoustic steps. 

%%%%%%%%%%%%%%%%%%%%%%%%%%%%%%%%%%%%%%%%%%%%%%%%%%%%%%%%%%%%%
% Section
%%%%%%%%%%%%%%%%%%%%%%%%%%%%%%%%%%%%%%%%%%%%%%%%%%%%%%%%%%%%%
\section{Numerical examples}\label{validation}
In this part, several benchmark tests, including standing wave, oscillation drop and dam-break flow are first carried out to validate the robustness, accuracy and energy conservation properties of the proposed method. Having the validation in hand, an ocean engineering application of wave interaction i.e. OWSC \cite{44} is simulated to investigate the versatility and performance. The ratio of smoothing length to the particle resolution i.e. $h/dp$ is 1.3 without specification and the corresponding cutoff radius is 2.6 $dp$. It should be noted that the water pressure is zero at the initial state.
%%%%%%%%%%%%%%%%%%%%%%%%%%%%%%%%%%%%%%%%%%%%%%%%%%%%%%%%%%%%%
% SubSection
%%%%%%%%%%%%%%%%%%%%%%%%%%%%%%%%%%%%%%%%%%%%%%%%%%%%%%%%%%%%%
\subsection{Standing wave}\label{Standing wave}
\begin{figure}[htb!]
	\centering
	\includegraphics[width=0.7\textwidth]{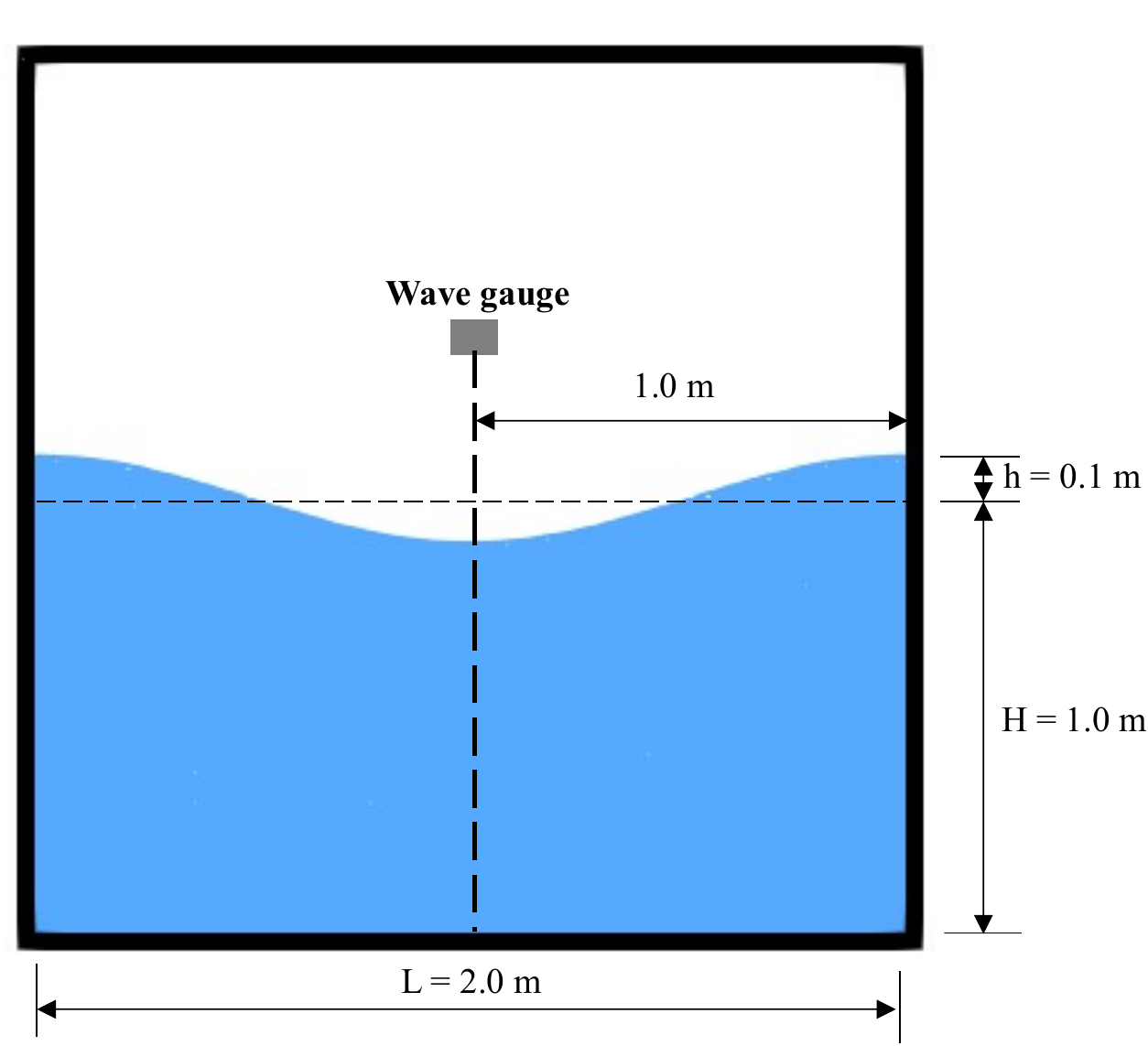} \\
	\caption{Standing wave: The sketch of the standing wave case.}
	\label{figs:sketch of standing wave}
\end{figure}

In this section, a 2D standing wave is computed with the initial sketch shown in Figure.\ref{figs:sketch of standing wave}. The initial free surface is given as
\begin{equation}
\eta_0=Acos\left(k(x+\lambda)/2\right),\label{eq22}
\end{equation}
where the wave amplitude $A = 0.1 H$, the wave number $k=2\pi / \lambda$ and the wave length $\lambda=2$ m. The average water depth is $H = 1.0$ m and the initial velocity of particles is zero. The free-surface elevation at the center position is measured and compared with the second-order analytical solution derived by Wu and Taylor \cite{26} and numerical results in Ref. \cite{25}.

\begin{figure}[htb!]
	\centering
	\includegraphics[width=1.0\textwidth]{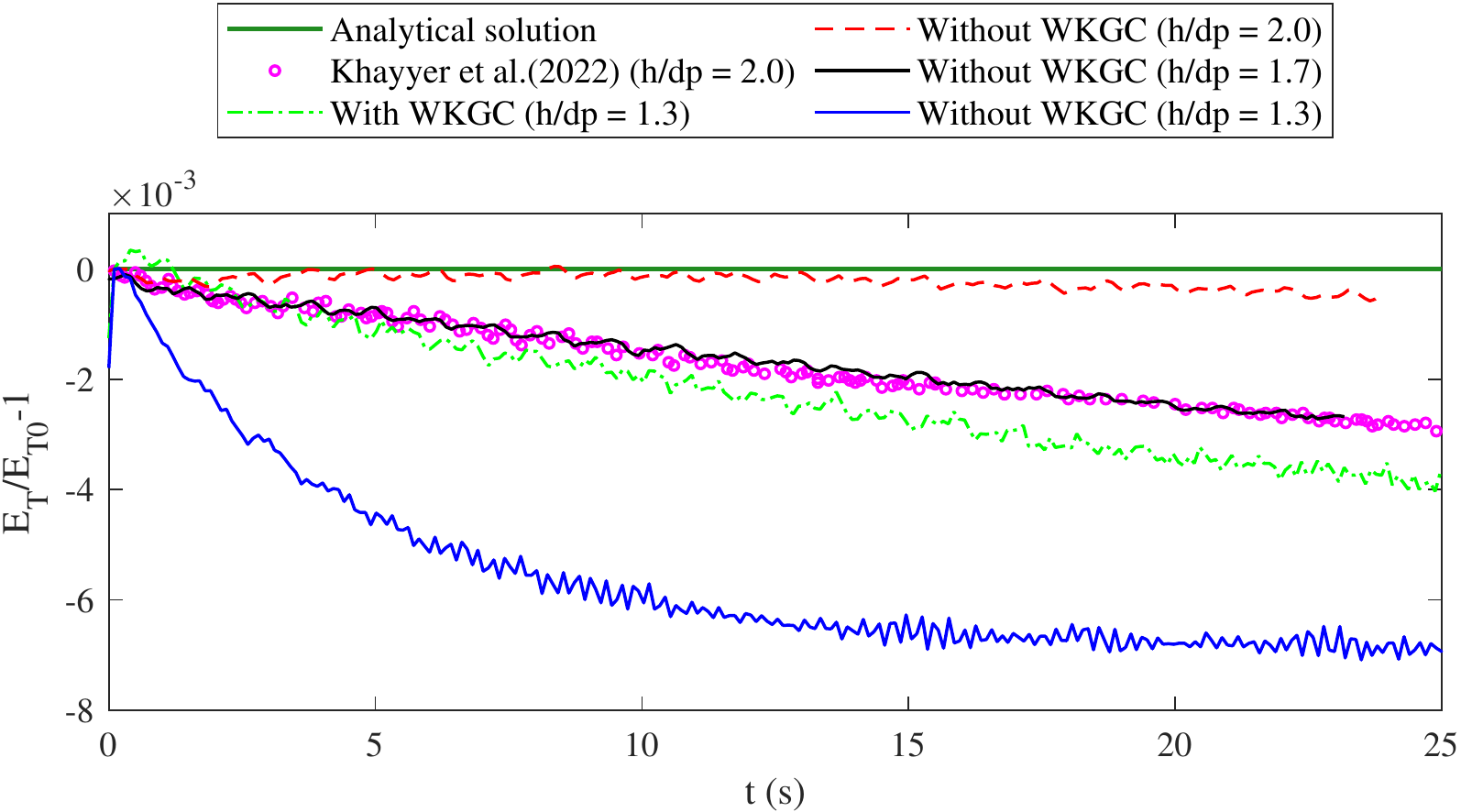} \\
	\caption{Standing wave: Time history of the normalized mechanical energy with different smoothing lengths. Comparison is conducted against the analytical solution and numerical results of Khayyer et al. \cite{25}($H/dp = 100$).}
	\label{figs:Standingwave-energy-compare}
\end{figure}

Figure.\ref{figs:Standingwave-energy-compare} shows the time variation of the normalized mechanical energy. As expected, the present method greatly reduces the numerical dissipation compared with those without WKGC. To study the effect of the smoothing length on the energy conservation property, the time history of the normalized mechanical energy predicted by the method without the WKGC scheme under different smoothing lengths is compared with those obtained by Khayyer et al.\cite{25} and an analytical solution, as also shown in Figure.\ref{figs:Standingwave-energy-compare}. It can be noted that increasing the smoothing length leads to a clear enhancement in the energy conservation properties but excessive computation efforts. More precisely, a good agreement with the analytical solution is noted for the simulation without KGC when $h = 2.0~dp$ is applied, while excessive numerical dissipation is exhibited in Ref. \cite{25} with a similar smoothing length setup. 

\begin{figure}[htb!]
	\centering
	\includegraphics[width=1.0\textwidth]{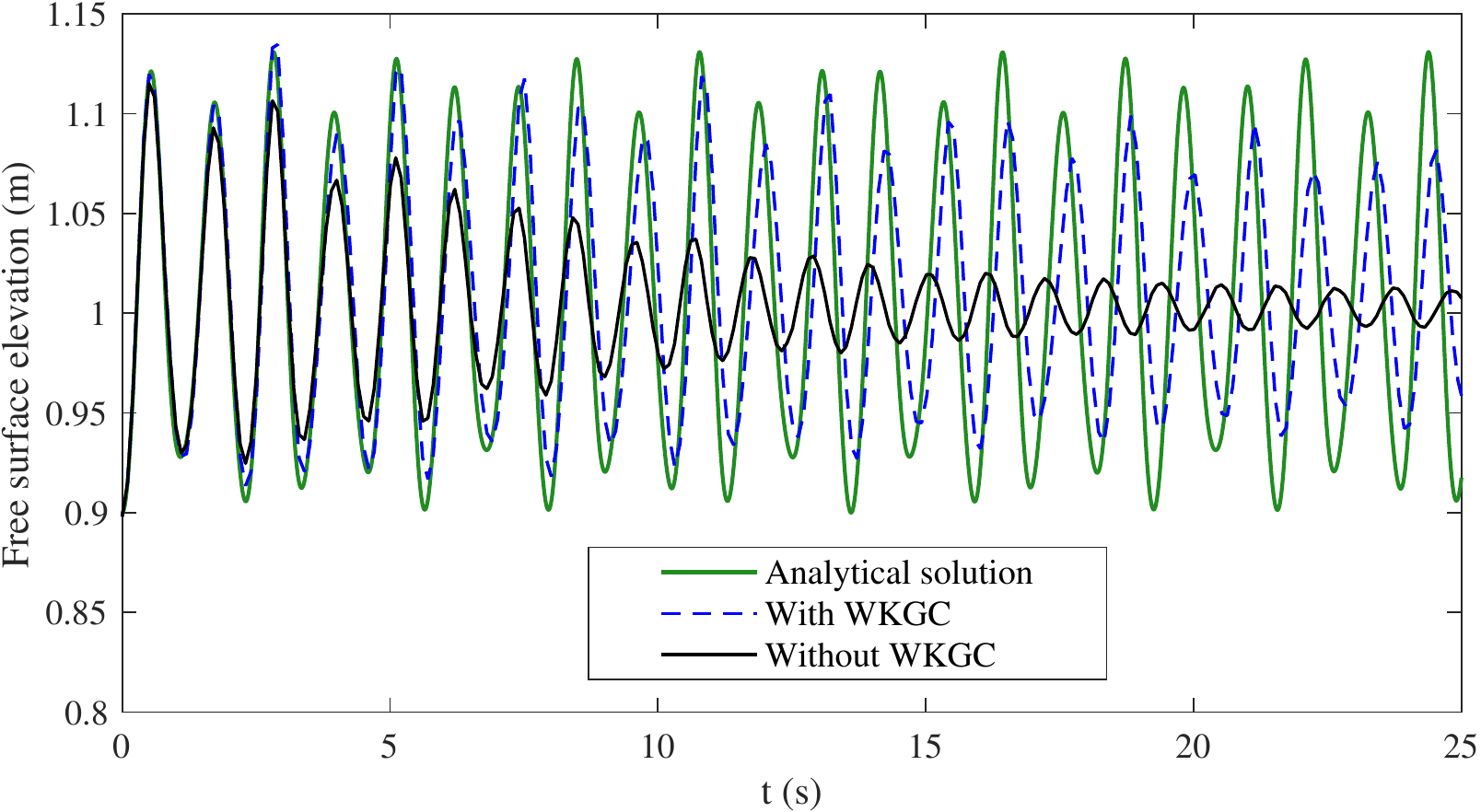} \\
	\includegraphics[width=1.0\textwidth]{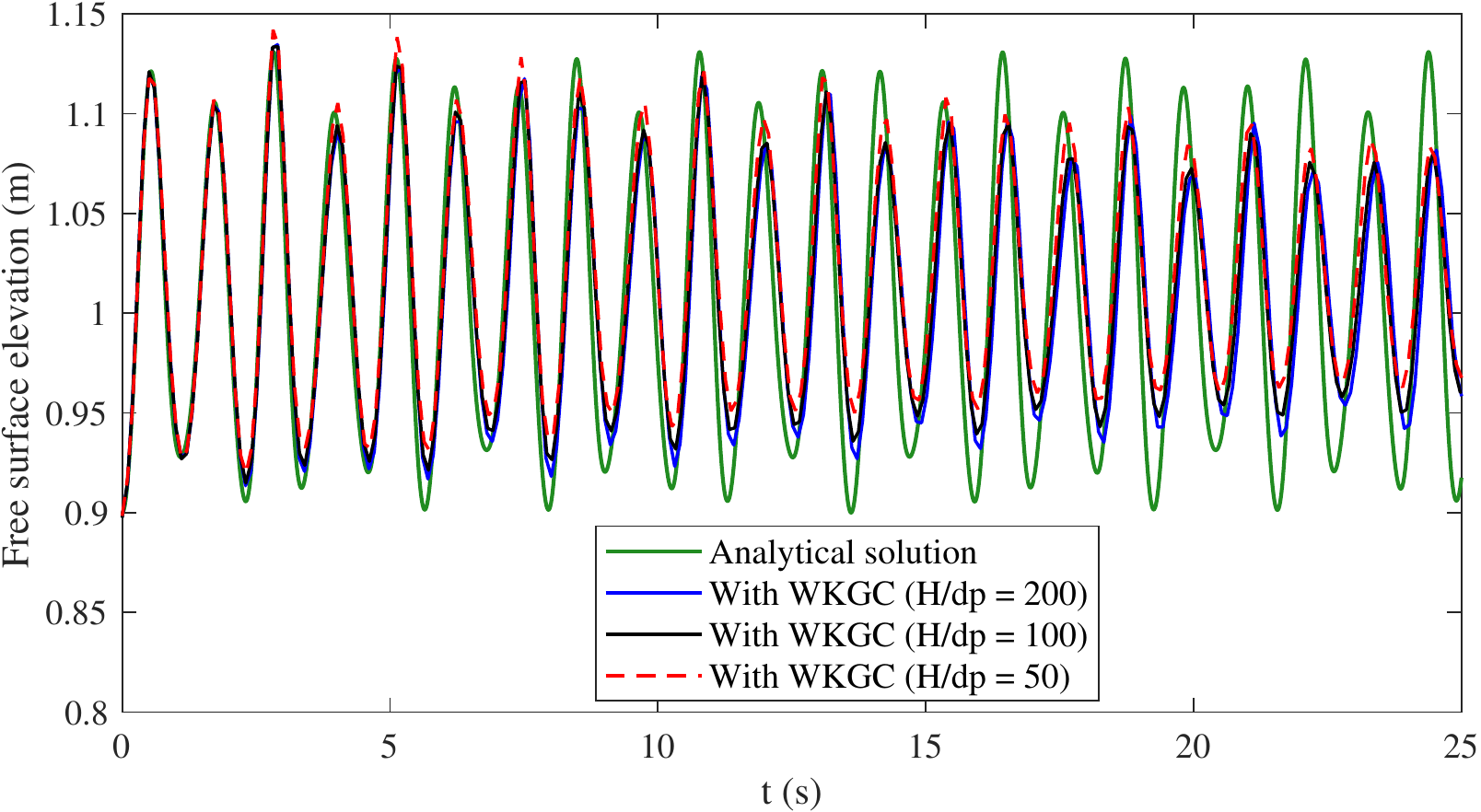} 
	\caption{Standing wave: Time evolution of the free-surface elevation at the center of the tank. (a) The comparison between the method with and without the WKGC scheme ($H/dp = 200$) and (b) the convergence study.}
	\label{figs:standingwave-elevation}
\end{figure}

Figure.\ref{figs:standingwave-elevation} (a) presents the time history of the free-surface elevation and its comparison with the analytical solution. As expected, the WKGC scheme significantly improves the numerical accuracy with fewer errors from the analytical peak and phase. Figure.\ref{figs:standingwave-elevation} (b) displays the convergence of the free-surface elevation with spatial resolutions. As the increase of particle resolution, convergence with the analytical solution is obtained.

\begin{figure}[htb!]
	\centering
	\includegraphics[width=1.0\textwidth]{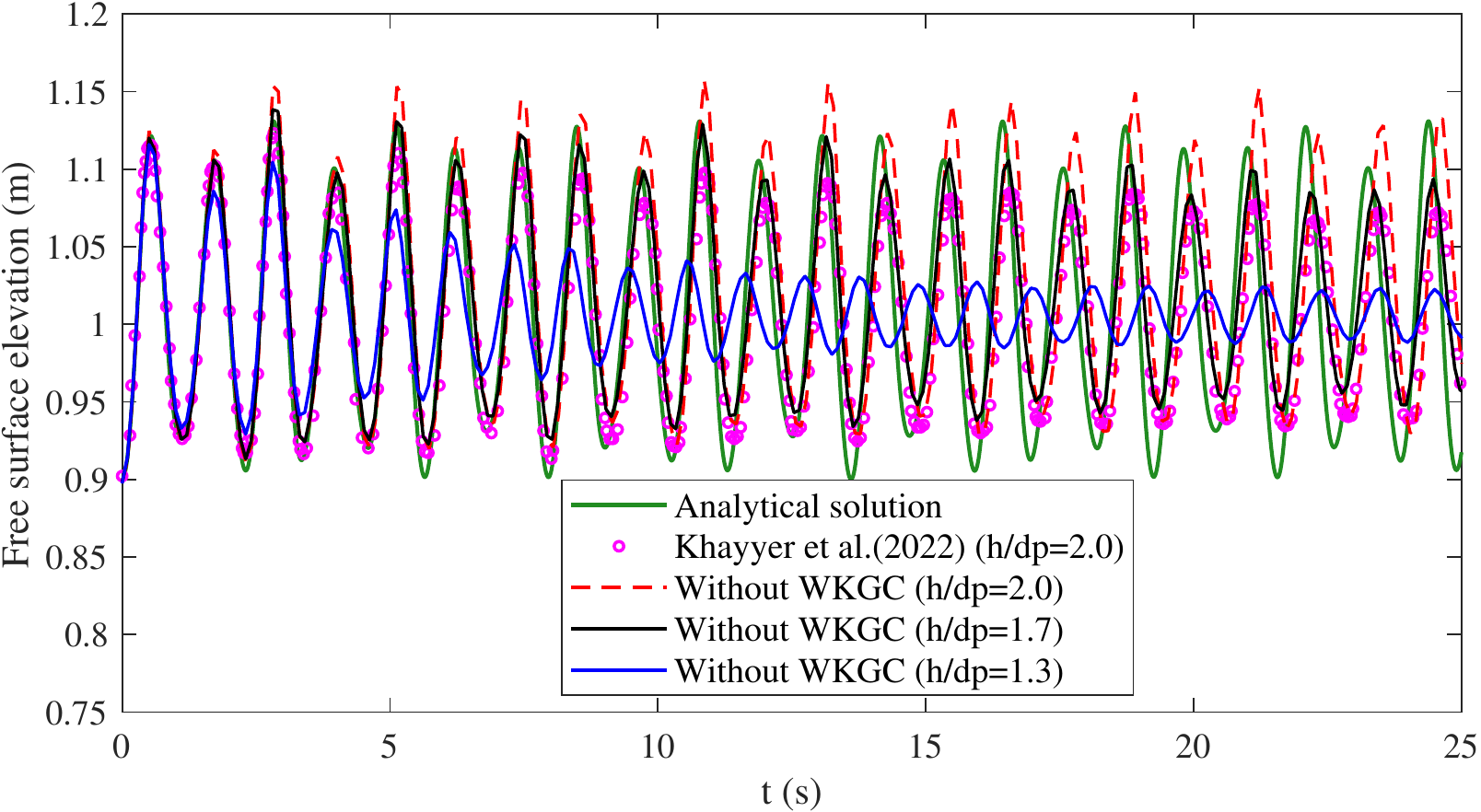} \\
	\caption{Standing wave: Time history of the free-surface elevation at the tank center by the method without WKGC for different smoothing lengths. Comparison is conducted against the analytical solution and numerical results of Khayyer et al.\cite{25}($H/dp = 100$).}
	\label{figs:Standingwave-elevation-Compared-Khayyer}
\end{figure}

Figure.\ref{figs:Standingwave-elevation-Compared-Khayyer} presents predicted free-surface elevations by the method without the WKGC scheme for different smoothing lengths. Similar to the profile of mechanical energy, increasing the smoothing length provides a clear convergence with the analytical solution. To investigate the extra computational efforts induced by increasing the smoothing length, Table.\ref{table: StandingWaveCPUTime} shows the wall-clock CPU time for simulations corresponding to different smoothing lengths with and without the WKGC scheme. All the simulations are performed on a laptop with an Intel Core i7-9750H. As expected, the CPU cost increases significantly with the increase in smoothing length and the use of WKGC. Also, it is not difficult to conclude that the CPU cost increases exponentially with smoothing length for 3D simulations, as discussed in Section \ref{OWSC}.    

\begin{table}[t]
  \centering
  \caption{Standing wave: The CPU time with different smoothing lengths. The computations are performed on a laptop with an Intel Core i709750H CPU.}
  \resizebox{\textwidth}{!}{
  \begin{tabularx}{1.2\textwidth}{p{5em}<{\centering}p{8em}<{\centering}p{7em}<{\centering}p{7em}<{\centering}p{7em}<{\centering}}
    \hline
    $h/dp$ & Computing time & Physical time & Particle number& WKGC\\
    \hline
   1.3 & 38.43 s & 1.0 s & 23264  & No\\
   1.7 & 43.72 s & 1.0 s & 23264  & No\\
   2.0 & 46.78 s & 1.0 s & 23264  & No\\
   1.3 & 43.05 s & 1.0 s & 23264  & Yes\\
   1.7 & 48.57 s & 1.0 s & 23264  & Yes\\
   2.0 & 49.55 s & 1.0 s & 23264  & Yes\\
   \hline
\end{tabularx}
}
\label{table: StandingWaveCPUTime}
\end{table}
%%%%%%%%%%%%%%%%%%%%%%%%%%%%%%%%%%%%%%%%%%%%%%%%%%%%%%%%%%%%%
% SubSection
%%%%%%%%%%%%%%%%%%%%%%%%%%%%%%%%%%%%%%%%%%%%%%%%%%%%%%%%%%%%%
\subsection{Oscillation drop}\label{Oscillation drop}
%
%%%%%%%%%%%%%%%%%%%%%%%%%%%%%%%%%%%%%%%%%%%%%%%%%%%%%%%%%%%%%
In this part, we consider an oscillating drop, which is a typical benchmark test that has been studied in the literature \cite{28,39,40}, to investigate the energy conservation property of the proposed method. Following Ref.\cite{39}, the drop radius is $R$ and the fluid is considered to be inviscid. Also, the drop is under a central conservative force $f= -\Omega^2 R$ and initialized with  a velocity profile
\begin{equation}
\left\{\begin{aligned}
u_0&=A_0x\\
v_0&=-A_0y ,\label{eq23}
\end{aligned}\right.
\end{equation}
where $A_0 = 1.0$ and $A_{0}/\Omega =1.0$. Note that an analytical solution is available \cite{41} for quantitative validation.

\begin{figure}[htb!]
	\centering
    \includegraphics[width=0.8\textwidth]{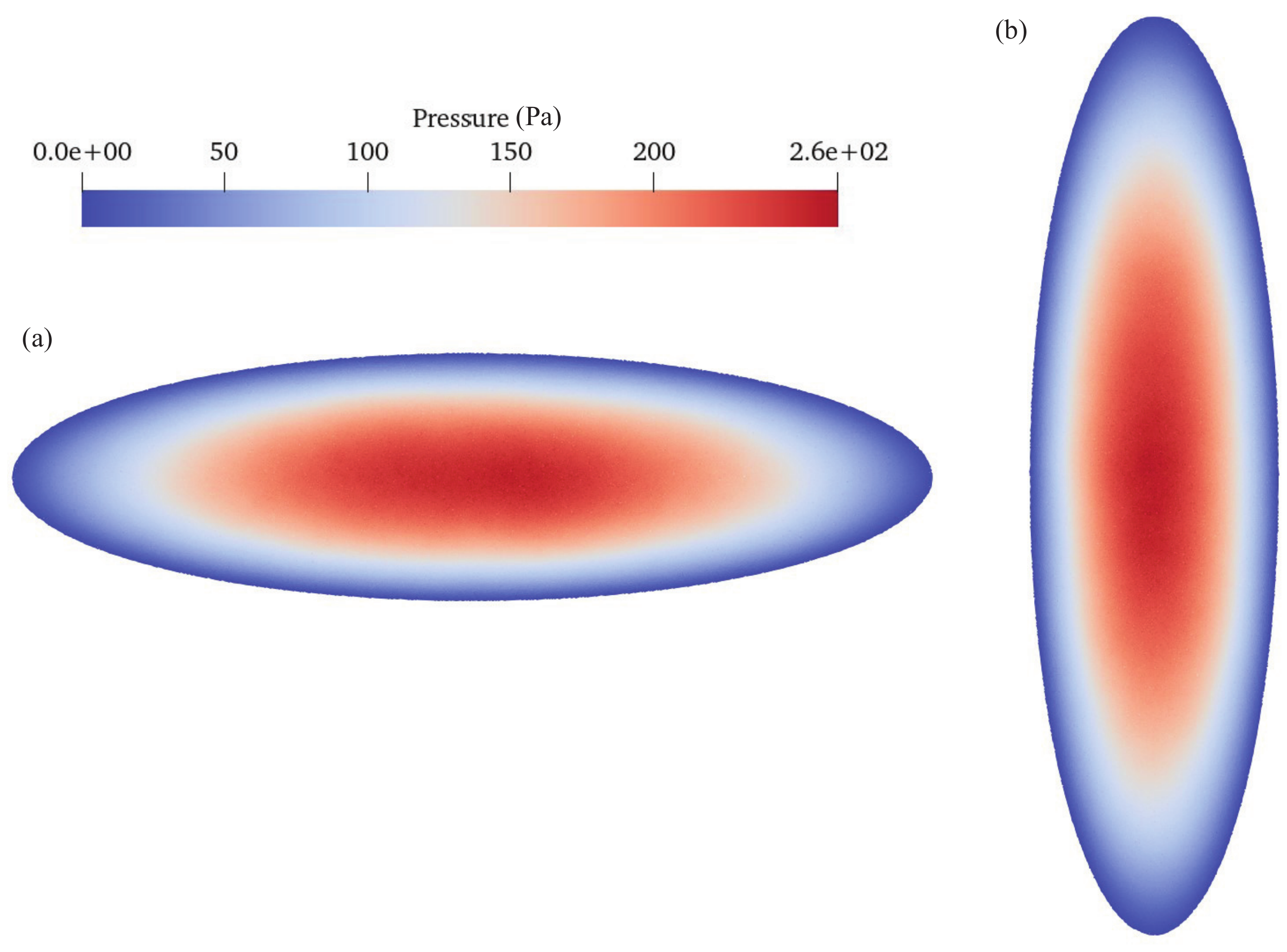}\\
	\caption{Oscillation drop: Snapshots of the free surface profile and the pressure contour reproduced by the present method (a) t = 20.5 s and (b) t = 22.9 s.}
	\label{figs:Oscillation drop pressure contour}
\end{figure}

\begin{figure}[htb!]
	\centering
	\includegraphics[width=1.0\textwidth]{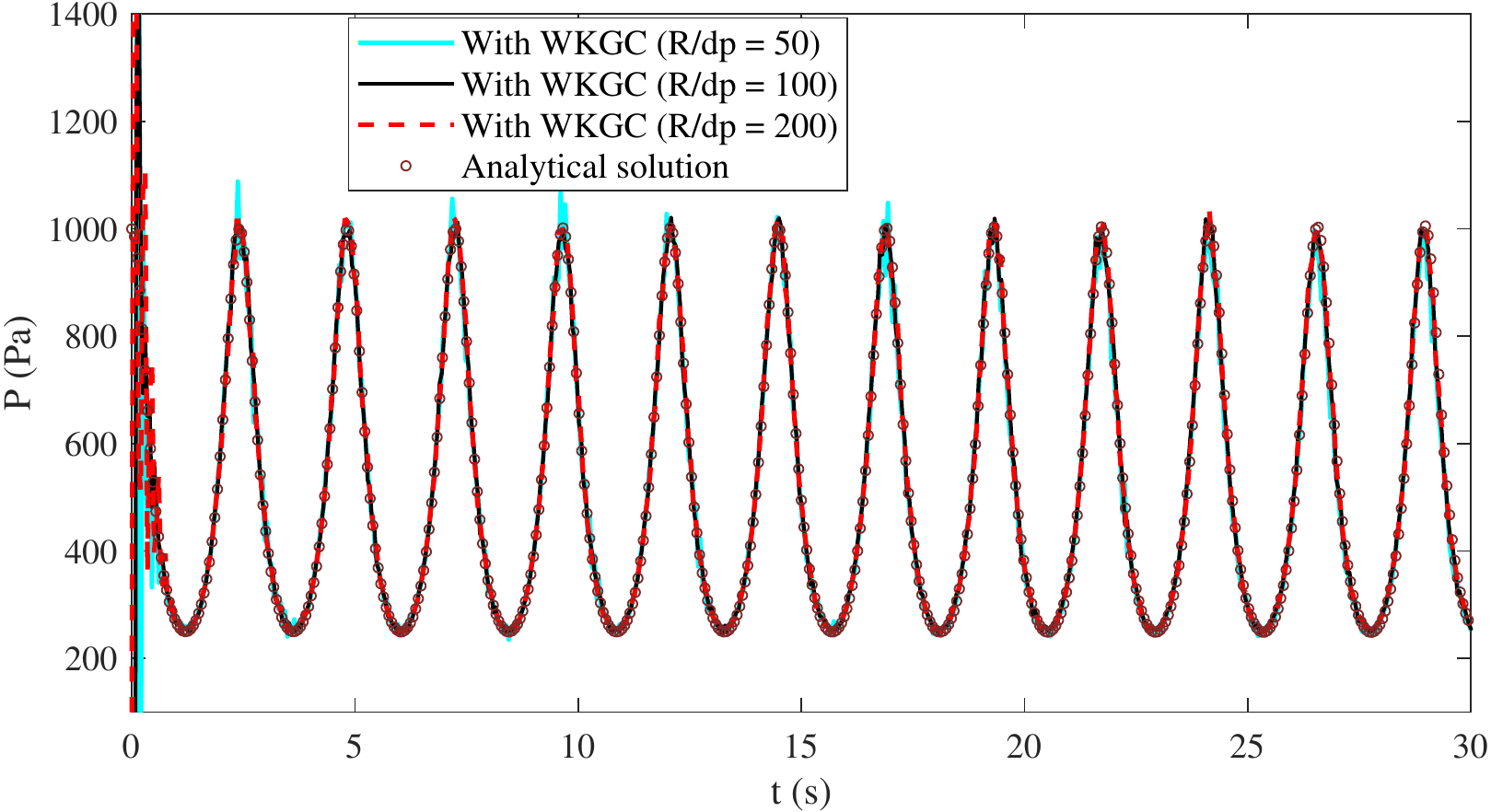} \\
	\caption{Oscillation drop: Time history of the pressure at the drop center with different particle resolutions.}
	\label{figs:Oscillation drop pressure sensor}
\end{figure}

\begin{figure}[htb!]
	\centering
	\includegraphics[width=1.0\textwidth]{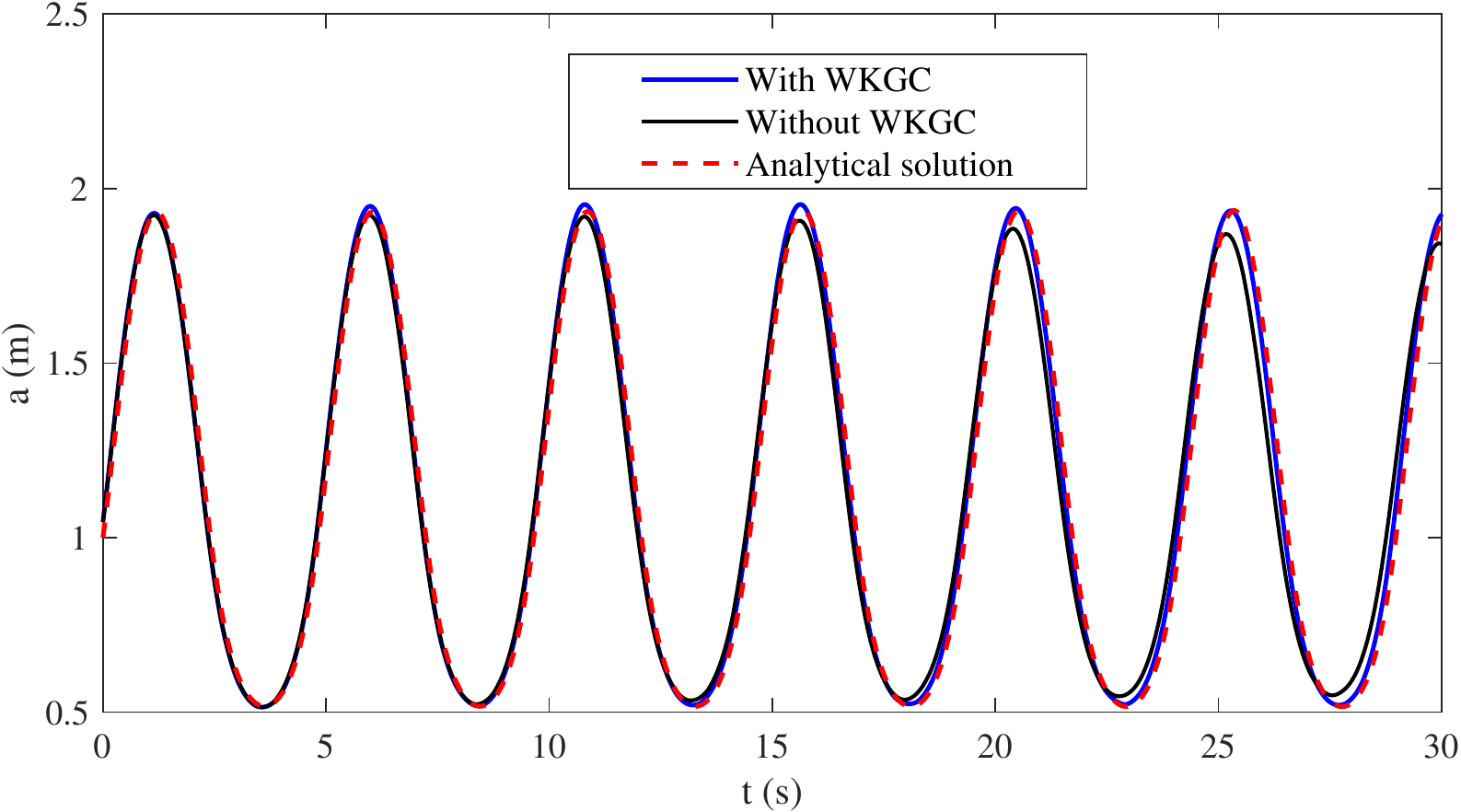} \\
	\caption{Oscillation drop: Time variation of the semi-major axis of the drop predicted by the method with and without the WKGC scheme ($R/dp=50$).}
	\label{figs:Oscillation drop axis}
\end{figure}

Figure.\ref{figs:Oscillation drop pressure contour} shows the free-surface profile and the pressure contour obtained by the present method at t = 20.5 s and t = 22.9 s. As expected, the present method produces a robust free-surface profile and smooth pressure field. The time variation of pressure at the drop center with different particle resolutions is presented in Figure.\ref{figs:Oscillation drop pressure sensor}. A good agreement with the analytical solution is noted with increasing spatial resolutions. Figure.\ref{figs:Oscillation drop axis} displays the time history of the semi-major axis of the drop. With the present WKGC scheme, higher numerical accuracy is achieved.

\begin{figure}[htb!]
	\centering
	\includegraphics[width=1.0\textwidth]{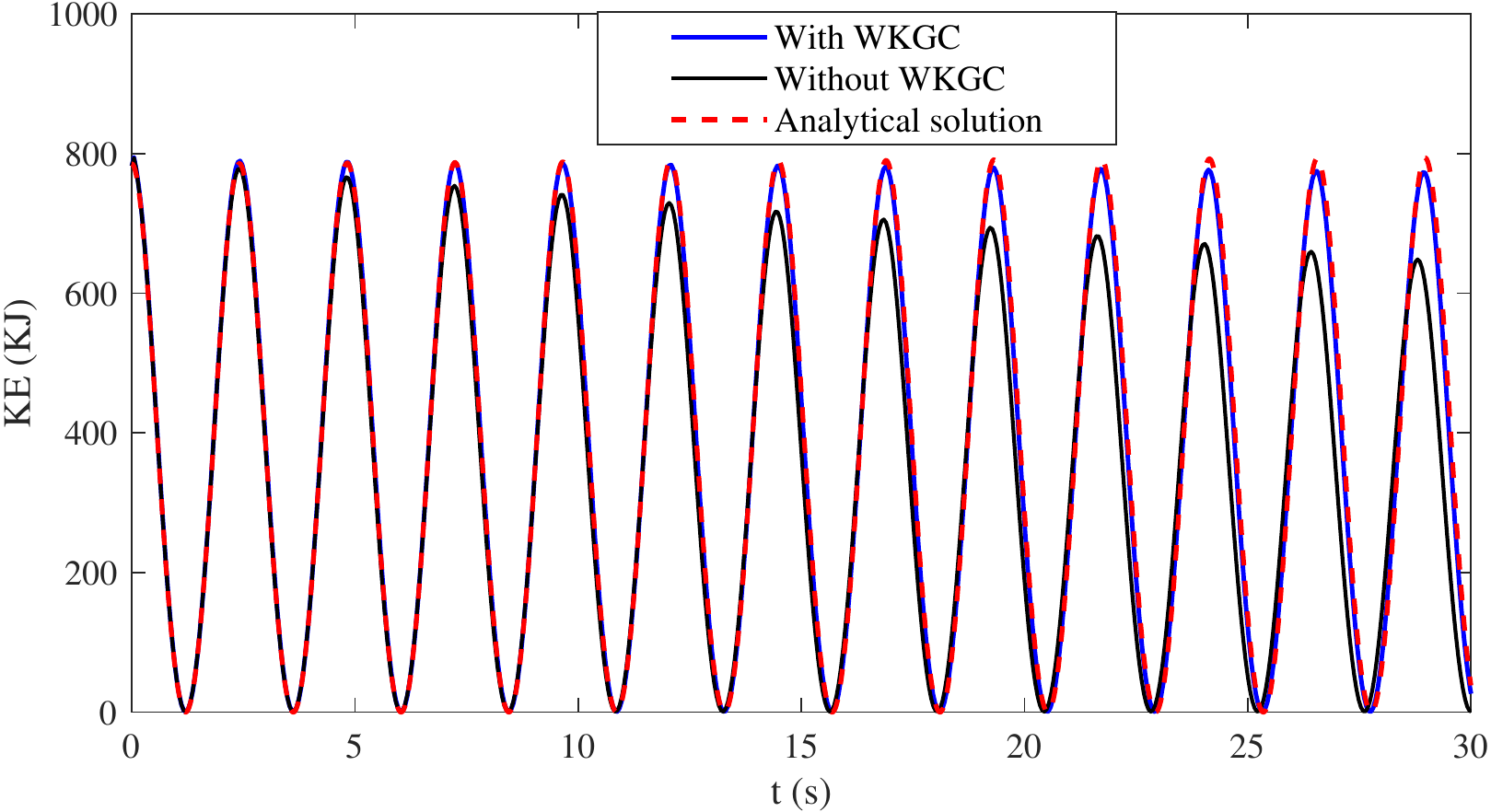} \\
	\includegraphics[width=1.0\textwidth]{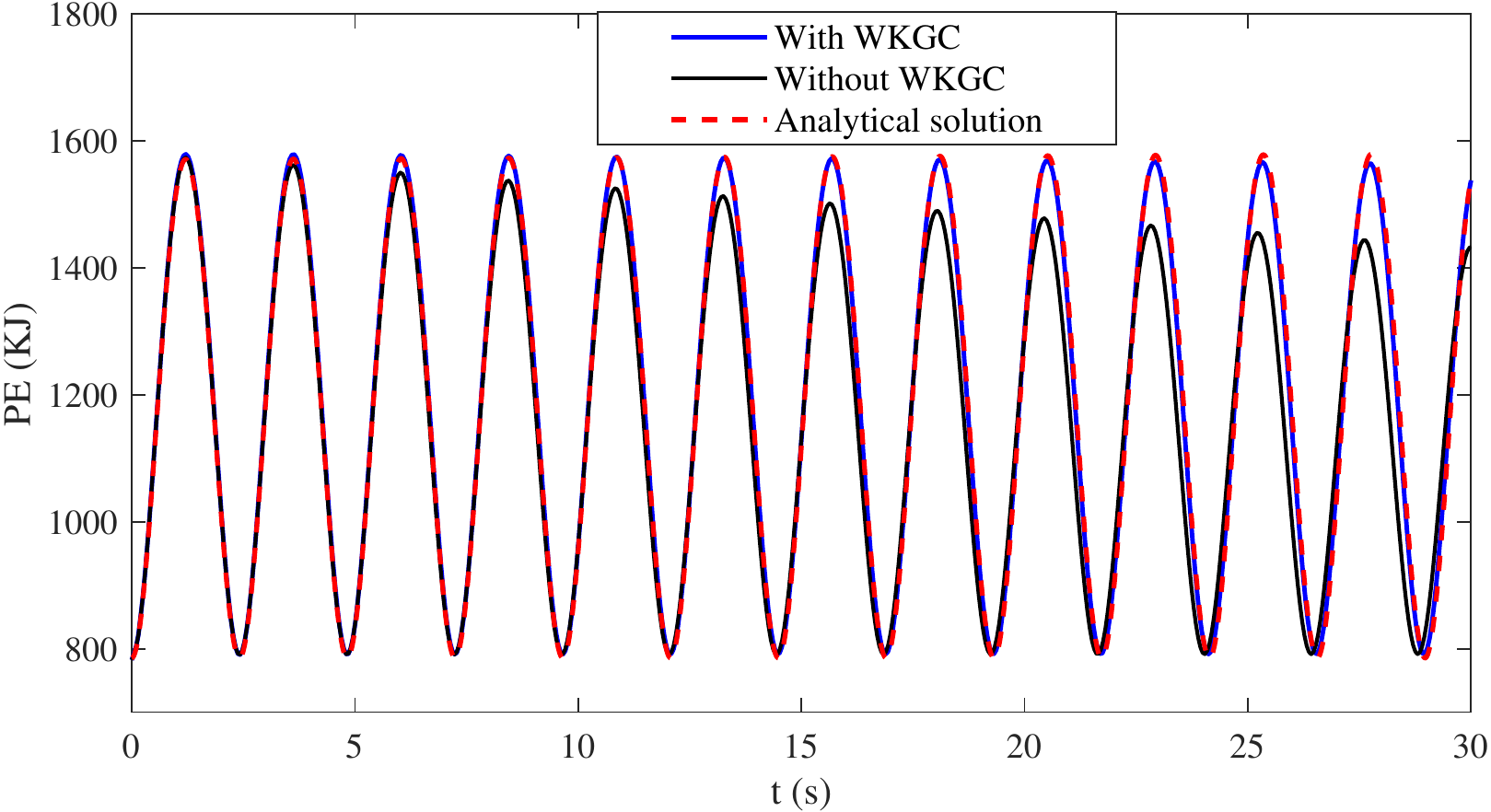} 
	\caption{Oscillation drop: Time history of the kinetic energy (a) and potential energy(b) by the method with and without the WKGC scheme ($R/dp=50$).}
	\label{figs:Oscillation drop kinetic and potential}
\end{figure}

\begin{figure}[htb!]
	\centering
	\includegraphics[width=1.0\textwidth]{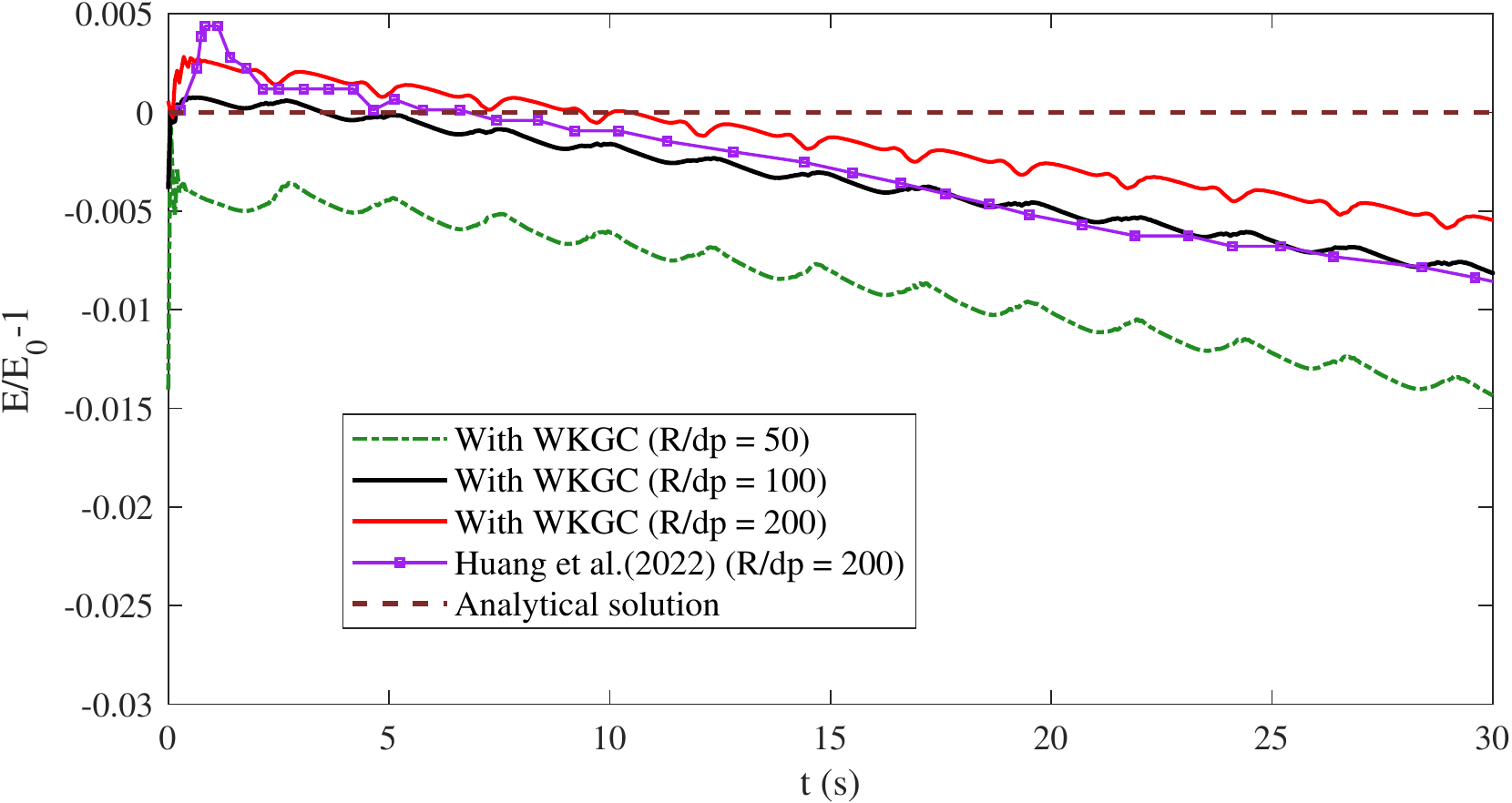} \\
	\caption{Oscillation drop: Time evolution of the normalized mechanical energy by the present method with different particle resolutions, Huang et al.\cite{38} and the analytical solution.}
	\label{figs:Oscillation drop mechniacal}
\end{figure}

\begin{figure}[htb!]
	\centering
	\includegraphics[width=1.0\textwidth]{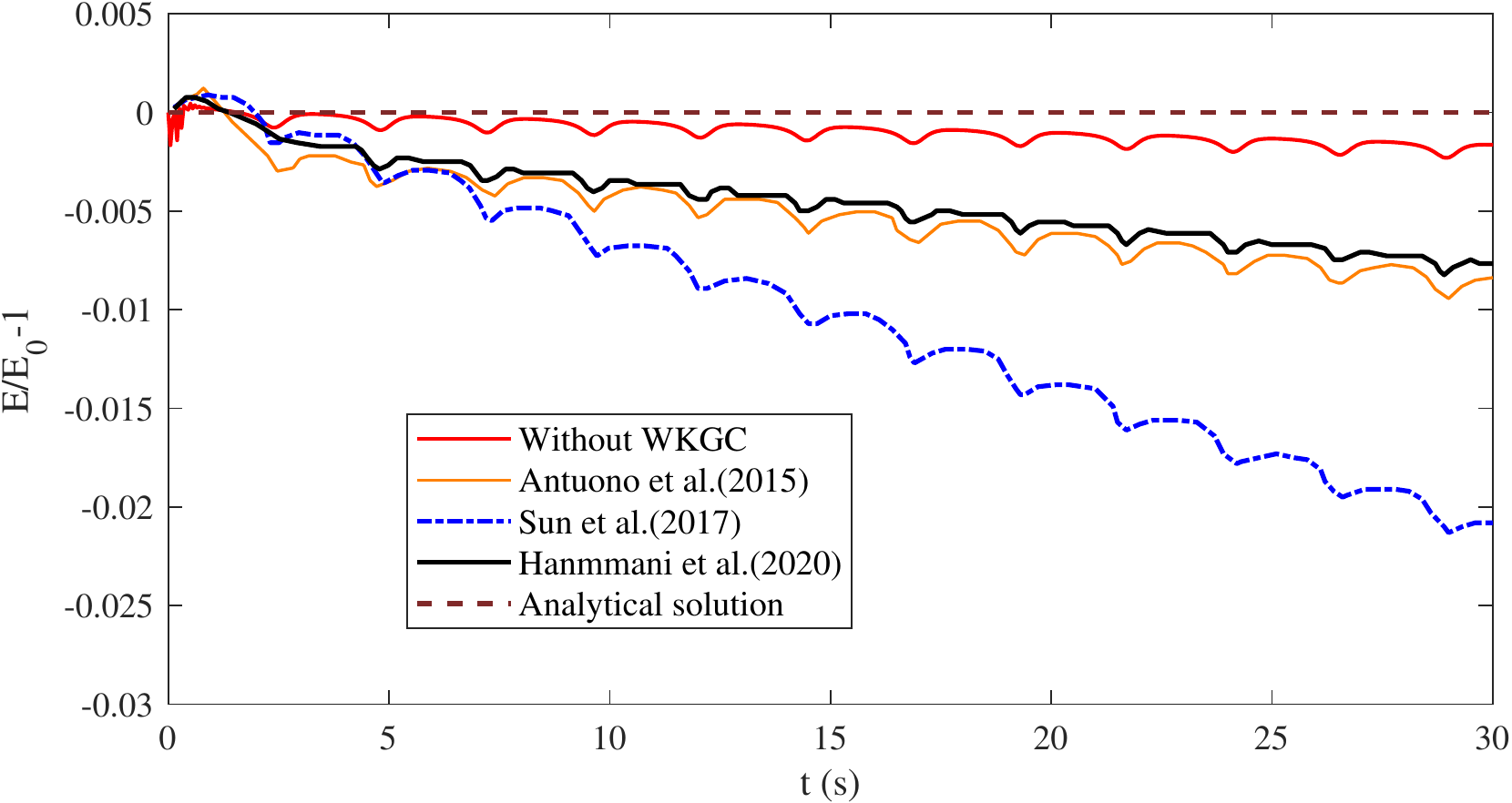} \\
	\caption{Oscillation drop: Time evolution of the normalized mechanical energy reproduced by the method without the WKGC scheme, Antuono et al.\cite{39}, Sun et al.\cite{28}, Hanmmani et al.\cite{40} and the analytical solution \cite{41} ($R/dp = 200$ and $ h/dp = 2.0$).}
	\label{figs:Oscillation drop mechniacal-2.0ds}
\end{figure}

To investigate the energy conservation property, the time variation of kinetic energy and potential energy is portrayed in Figure.\ref{figs:Oscillation drop kinetic and potential}. The present method presents good energy conservation characteristics and provides good consistency with the analytical solution. Figure.\ref{figs:Oscillation drop mechniacal} shows the time evolution of the normalized mechanical energy. Similar to Ref.\cite{38}, the mechanical energy first increases slightly and then decreases with time going. As reported by Colagrossi et al. \cite{59} and Huang et al. \cite{38}, the initial increase of the mechanical energy is due to the particle reorder. It is also observed that the present method exhibits less numerical dissipation compared with that of Huang et al. \cite{38}. Moreover, the mechanical energy rapidly convergences to the analytical solution as the particle resolution increases. Figure.\ref{figs:Oscillation drop mechniacal-2.0ds} shows the time evolution of the normalized mechanical energy by the method without WKGC, Antuono et al. \cite{39}, Sun et al. \cite{28}, Hanmmani et al. \cite{40} with a large smoothing length $h/dp = 2.0$. Without any KGC, the method with a large smoothing length achieves better energy conservation than the methods reported in the literature \cite{39,28,40}.


%%%%%%%%%%%%%%%%%%%%%%%%%%%%%%%%%%%%%%%%%%%%%%%%%%%%%%%%%%%%%
% SubSection
%%%%%%%%%%%%%%%%%%%%%%%%%%%%%%%%%%%%%%%%%%%%%%%%%%%%%%%%%%%%%
\subsection{Dam break}\label{dambreak}
\begin{figure}[htb!]
	\centering
	\includegraphics[width=1.0\textwidth]{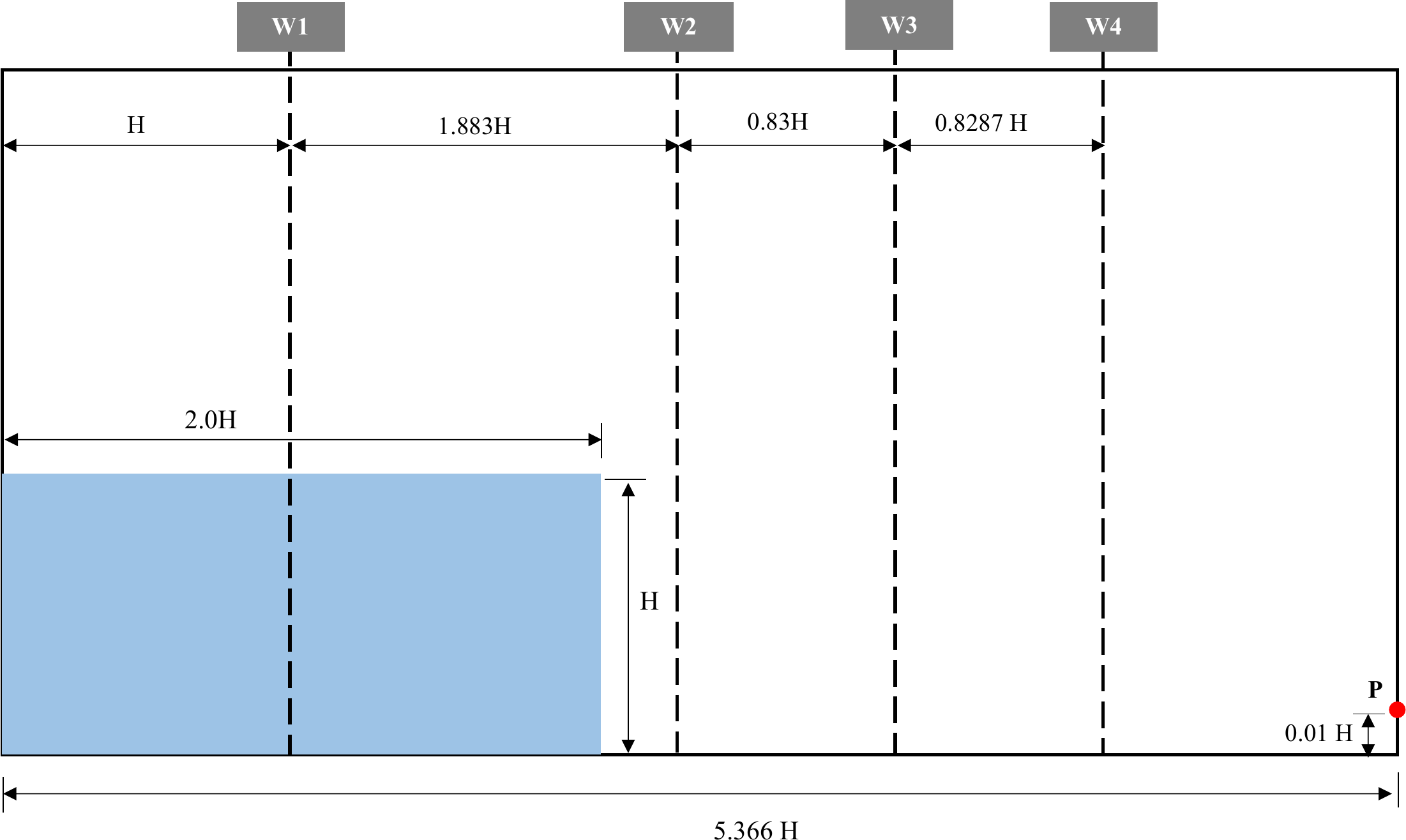}
	\caption{Dam-break: The sketch of the initial condition.}
	\label{figs:sketch of dambreak}
\end{figure}

The dam-break flow, which has been numerically \cite{48,49,50} and experimentally \cite{45,46,47} investigated in the literature, is a challenging case to evaluate the stability of a SPH algorithm. The schematic is depicted in Figure.\ref{figs:sketch of dambreak} where four probes (E1, E2, E3 and E4) and one pressure sensor (P) are set to measure the free-surface elevation and impact pressure on the wall, respectively, for quantitative validation. 

\begin{figure}[htb!]
	\centering
	\includegraphics[width=1.0\textwidth]{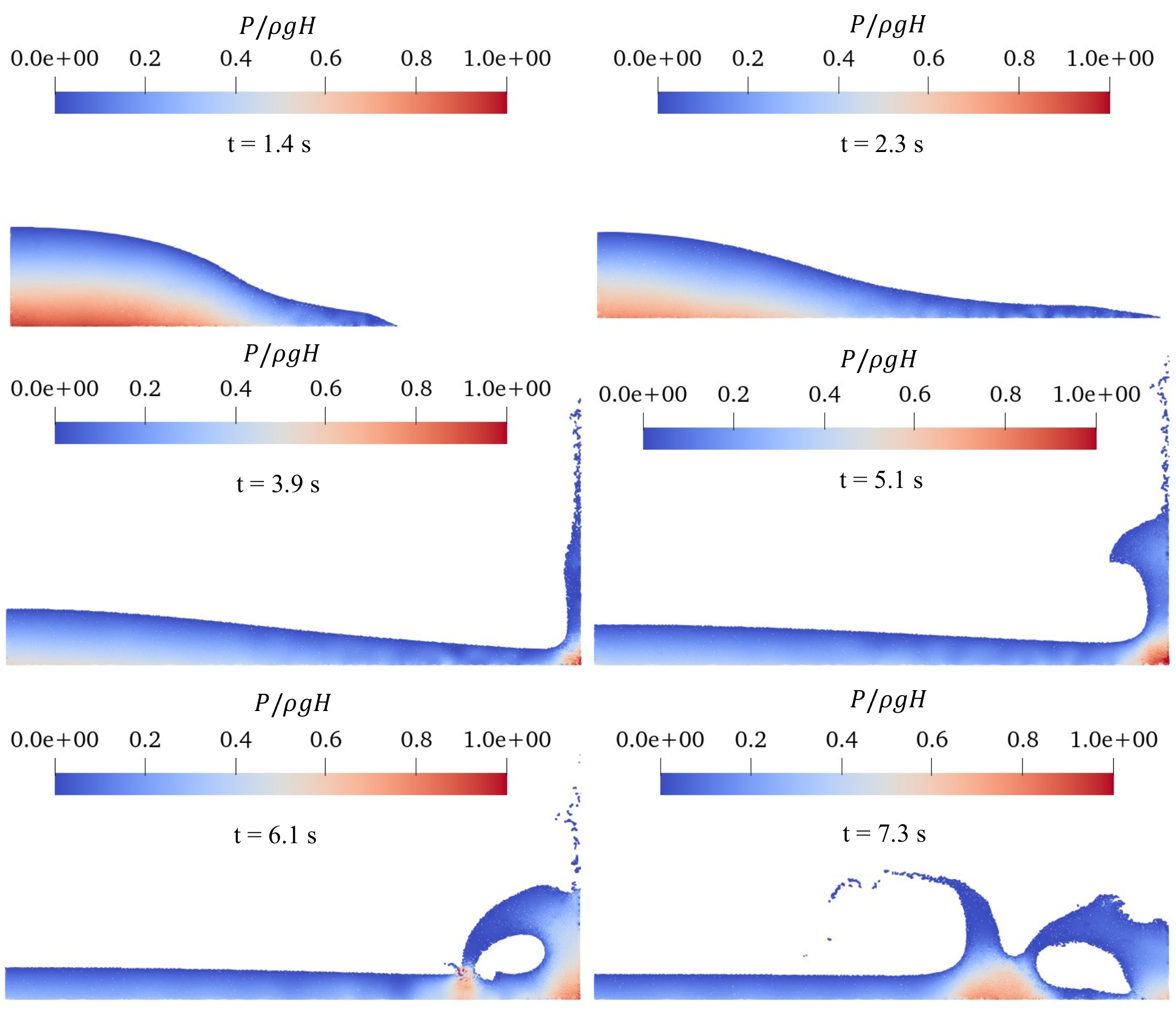} \\
	\caption{Dam-break: Snapshots of the free-surface profile and pressure contour reproduced by present method at different time instants.}
	\label{figs:Pressure}
\end{figure}

\begin{figure}[htb!]
	\centering
	\includegraphics[width=1.0\textwidth]{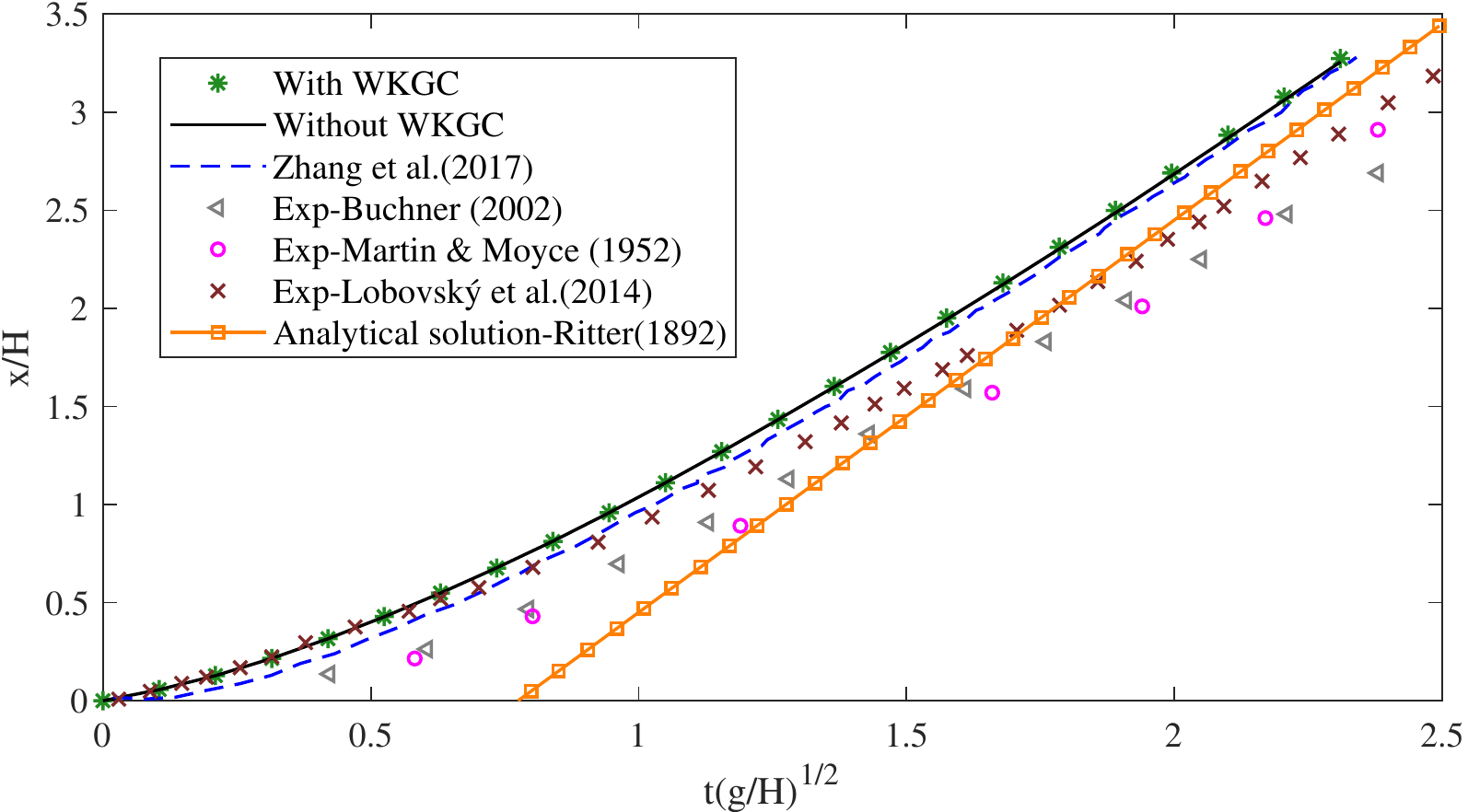}
	\caption{Dam-break: Time variation of the wavefront ($H/dp = 100$).}
	\label{figs:WaveFront}
\end{figure}

Figure.\ref{figs:Pressure} shows the snapshots of the free-surface profile and pressure contour at different time instants. The present method provides robust free-surface profiles and smooth pressure fields. The propagation wavefront predicted by the present method is compared with experimental data \cite{45,46,47} and the analytical solution derived by Ritter \cite{51} as shown in Figure.\ref{figs:WaveFront}. It can be observed that the present results agree well with experimental results \cite{46} before $t\sqrt{g/H}<1.0$, and are gradually close to the analytical solution at a later stage i.e. $t\sqrt{g/H}>1.0$. 

\begin{figure}[htb!]
	\centering
	\includegraphics[width=0.7\textwidth]{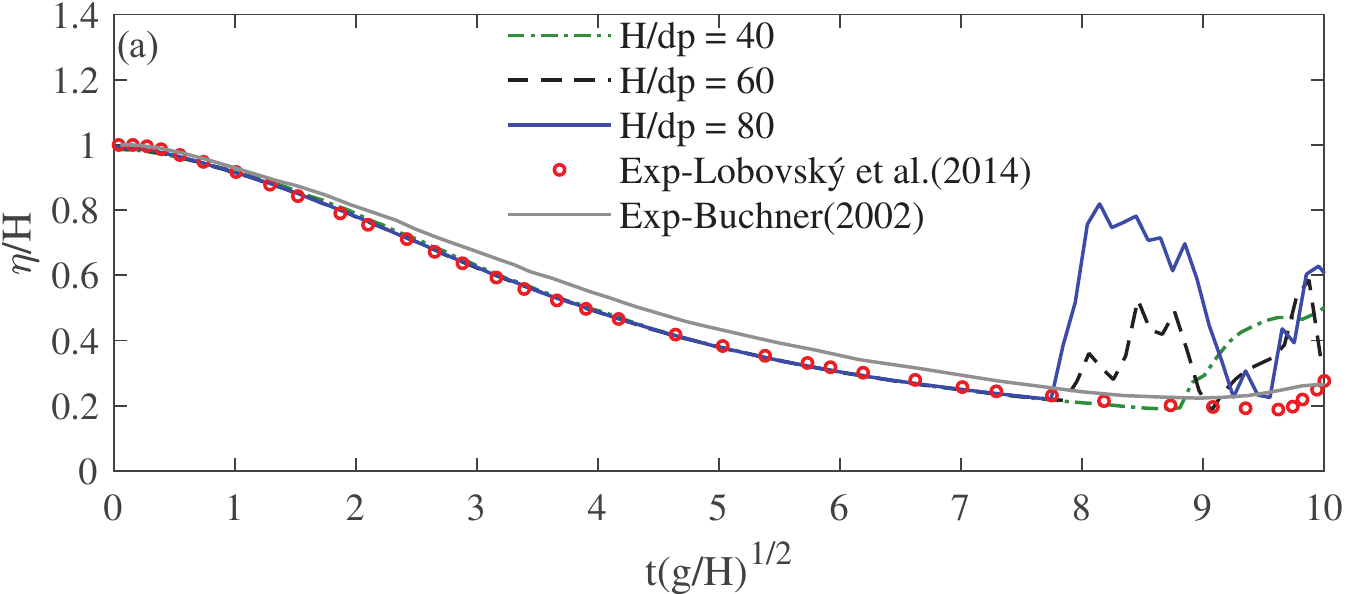} \\
	\includegraphics[width=0.7\textwidth]{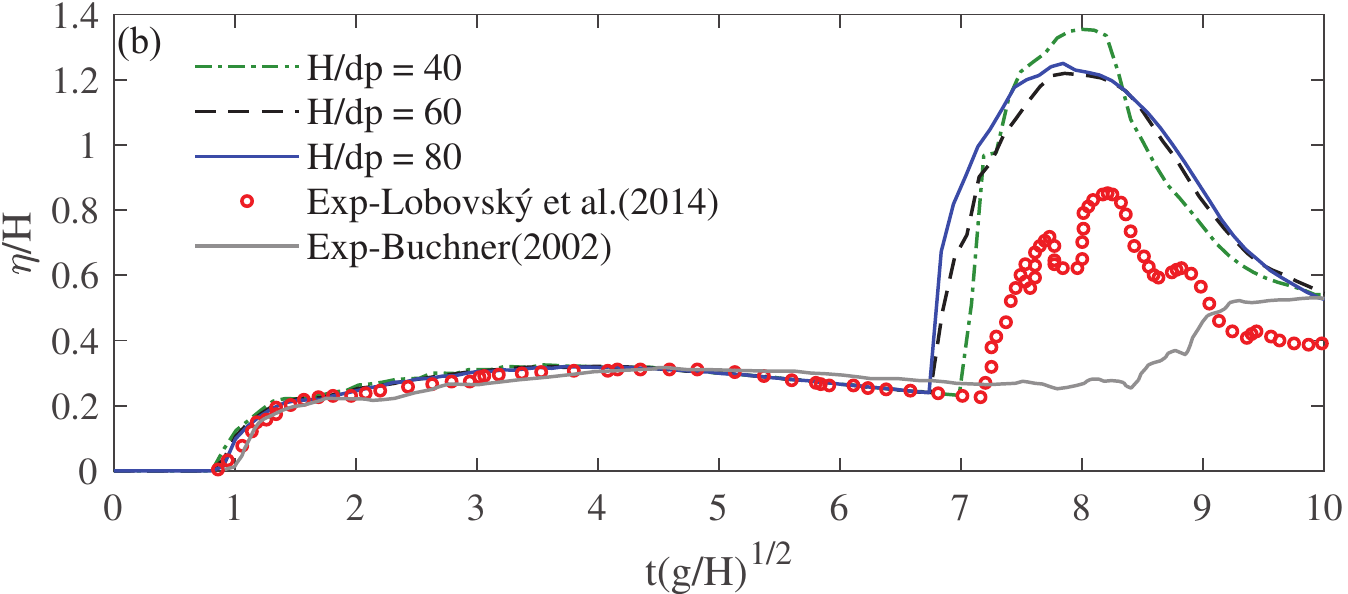} \\
       \includegraphics[width=0.7\textwidth]{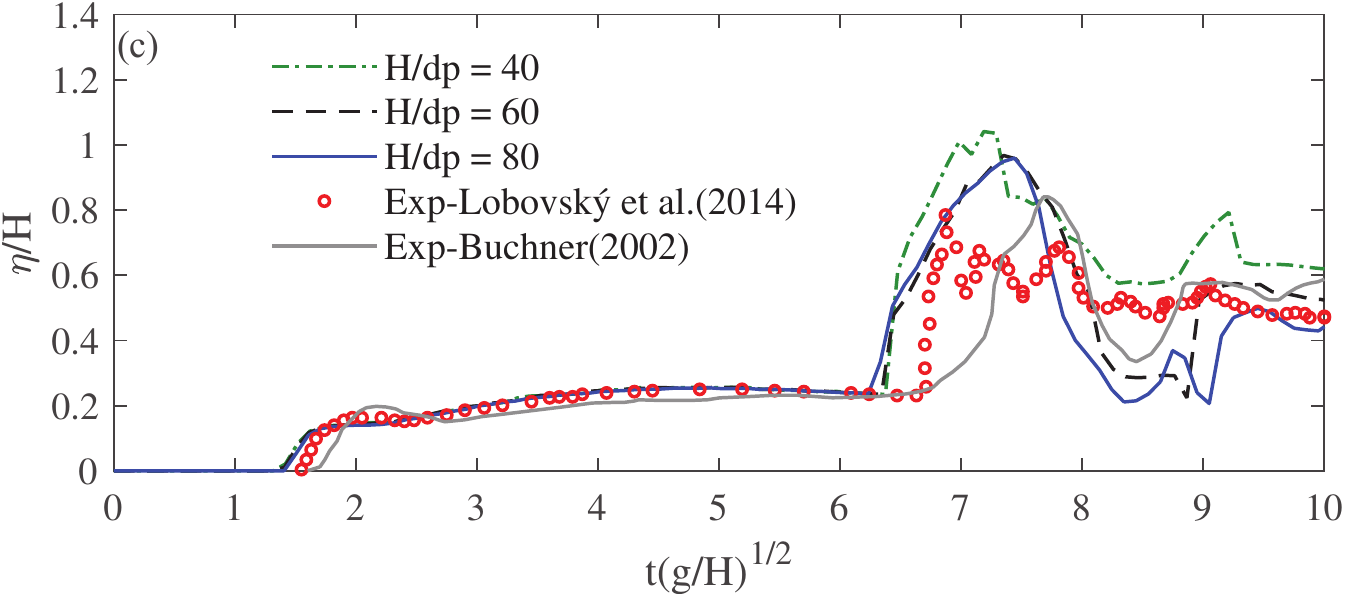} \\
       \includegraphics[width=0.7\textwidth]{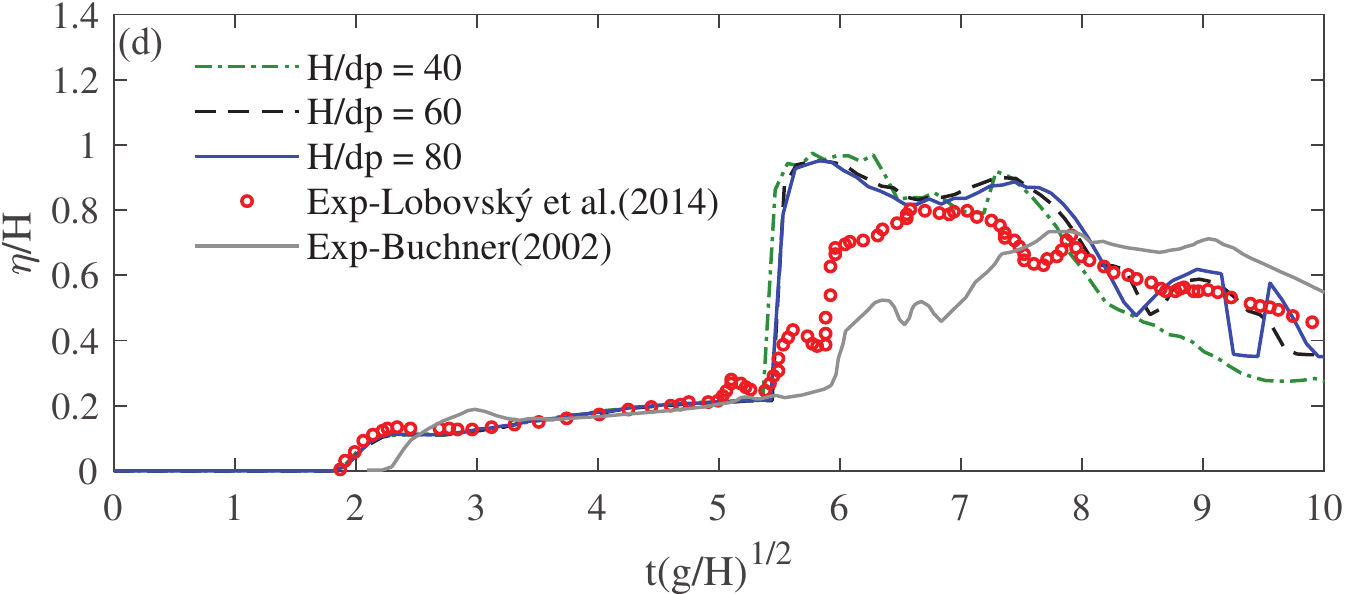} \\
       \caption{Dam-break: Time history of free-surface elevation at wave gauges W1 (a), W2 (b), W3 (c) and W4 (d).}
	\label{figs:Elevation}
\end{figure}

\begin{figure}[htb!]
	\centering
	\includegraphics[width=1.0\textwidth]{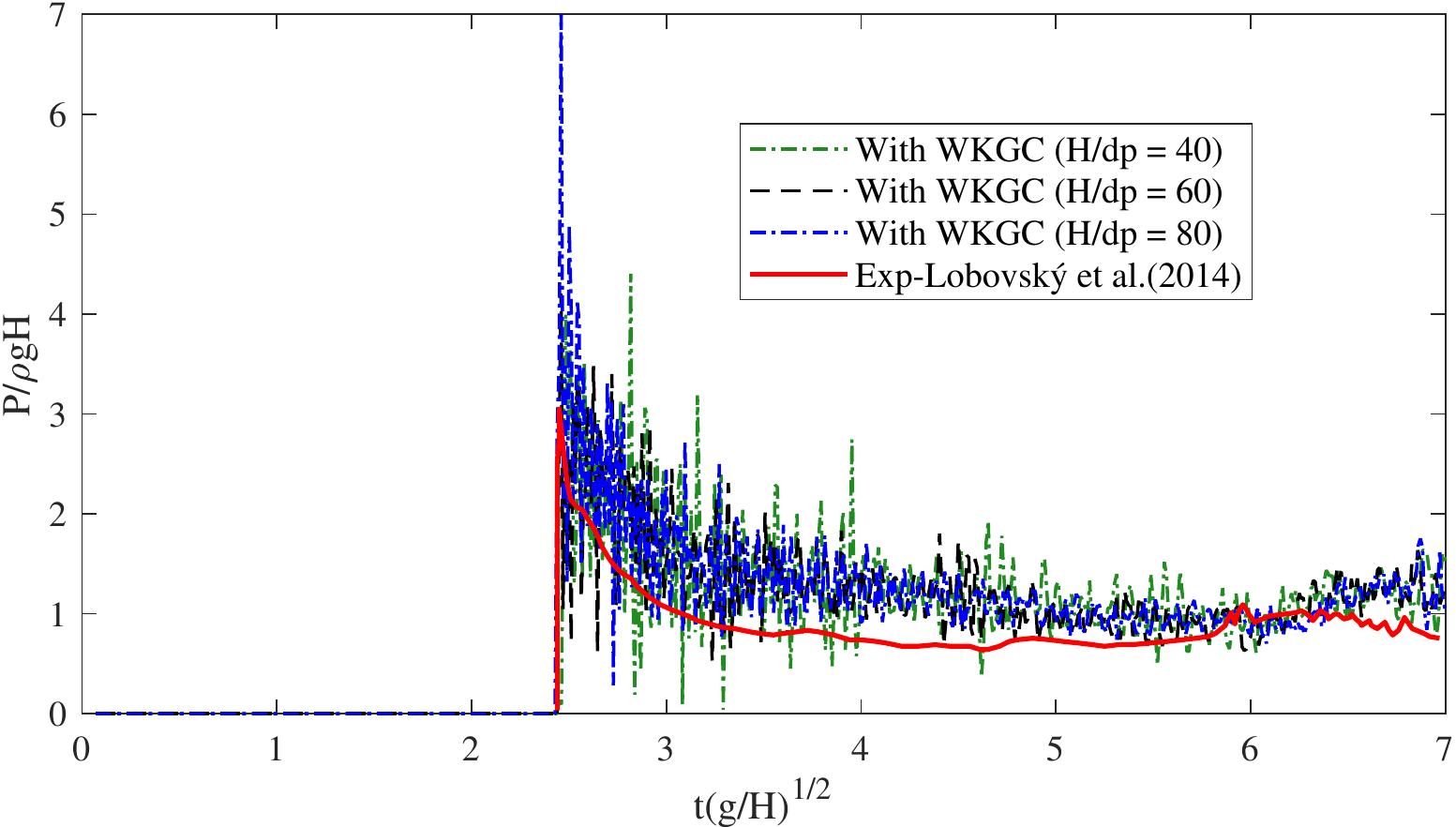} \\
	\caption{Dam-break: Time variation of the pressure probed at P.}
	\label{figs:Pressuresensor}
\end{figure}

Figure.\ref{figs:Elevation} shows the time history of the free-surface elevation. The numerical results have good consistency with experimental data obtained by Lobovsk`y et al. \cite{46}. Compared with the experimental data reported by Buchner \cite{45}, the present dam break waves propagate faster similar to the observation of Lobovsk`y et al. \cite{46}, leading to higher-up waves. Lobovsk`y et al. \cite{46} pointed out that this may be due to the fact that the bed downstream in his experiment was completely dry, while not in the experiment of  Buchner \cite{45}. It should be noted that there are some discrepancies in the arrival time and the following evolution of the secondary wave \cite{46} between the numerical results and experimental observation. This is mainly due to the fact that the secondary wave accompanies the wave breaking and re-entry. The time history of the pressure probed at P is presented in Fig.\ref{figs:Pressuresensor}. The numerical results are consistent with previous simulations reported in Refs \cite{50,48} and are in reasonable agreement with experimental data, except for the pressure oscillation resulting from the weakly compressible assumption.

\begin{figure}[htb!]
	\centering
	\includegraphics[width=1.0\textwidth]{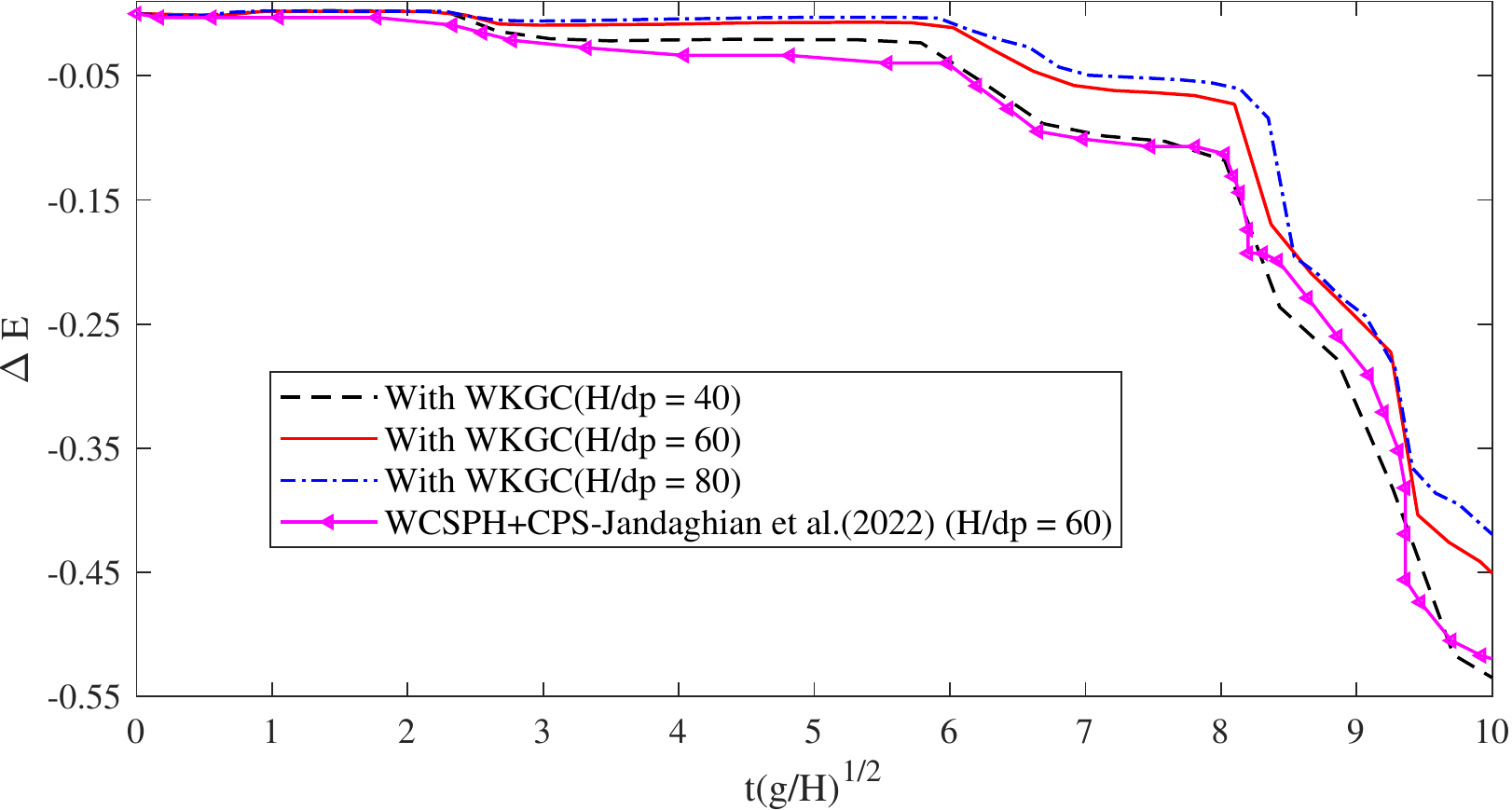} \\
	\includegraphics[width=1.0\textwidth]{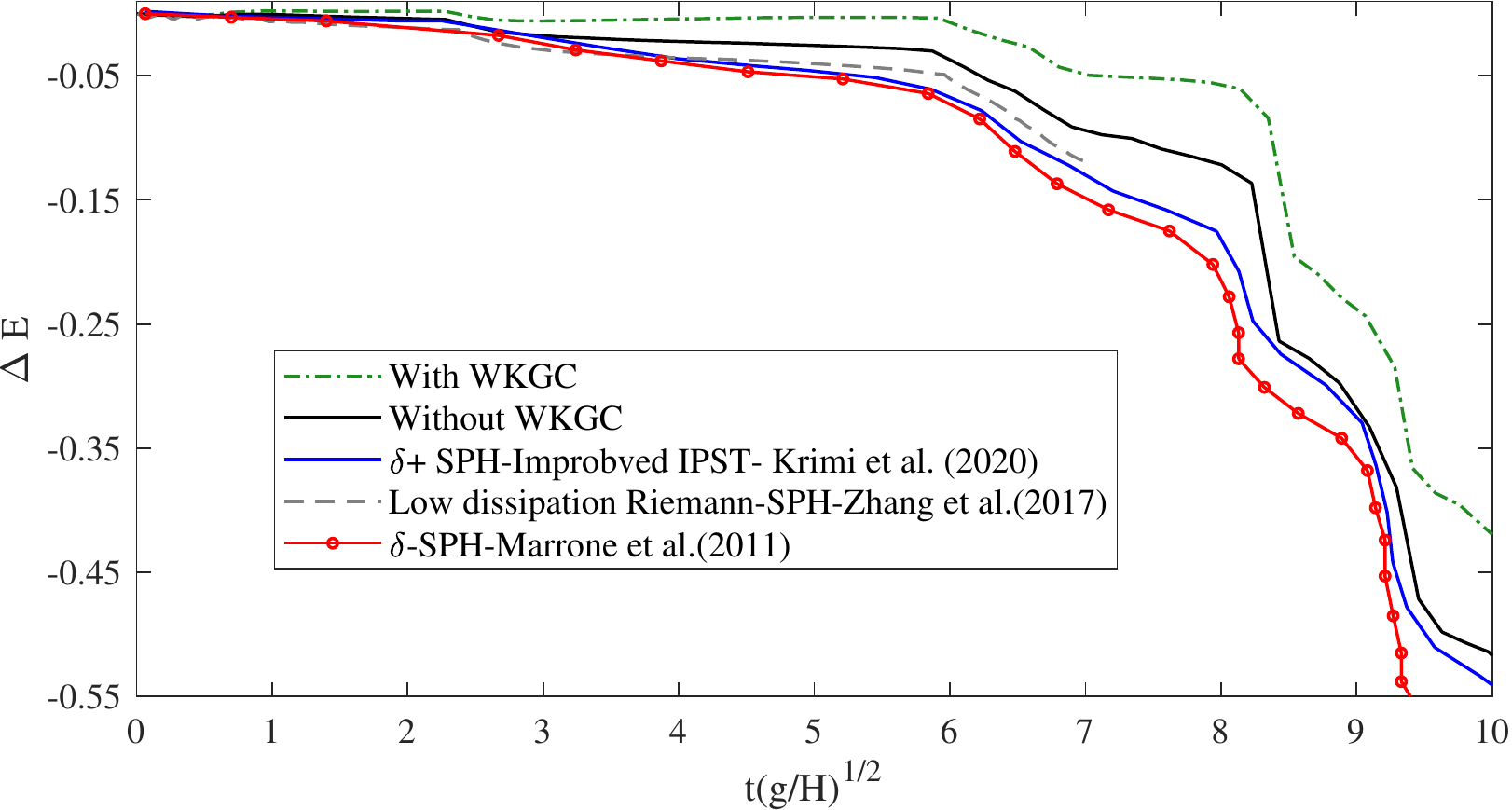} 
	\caption{Dam-break: Time evolution of the mechanical energy. Convergence study of the present method (a) and the comparison with other numerical results in the literature ($H/dp = 80$)(b).	}
	\label{figs:dambreak energy}
\end{figure}

Figure.\ref{figs:dambreak energy} displays the mechanical energy evolution and its comparison against numerical results in the literature \cite{50,51}. Following Refs.\cite{21,50}, the mechanical energy is normalized $(E-E_0)/(E_0-E_\infty),$ where E the mechanical energy, $E_0$ the initial mechanical energy and $E_\infty$ the mechanical energy after reaching the equilibrium state. As the particle resolution increases, the numerical dissipation rapidly decreases as shown in Figure.\ref{figs:dambreak energy} (a), indicating the convergence of the present method. Figure.\ref{figs:dambreak energy} (b) shows that the present method outperforms the method without WKGC in terms of energy conservation property. Also, the present method exhibits considerably less numerical dissipation compared with the results in the literature.

%%%%%%%%%%%%%%%%%%%%%%%%%%%%%%%%%%%%%%%%%%%%%%%%%%%%%%%%%%%%%
% SubSection
%%%%%%%%%%%%%%%%%%%%%%%%%%%%%%%%%%%%%%%%%%%%%%%%%%%%%%%%%%%%%
\subsection{OWSC}
\label{OWSC}
\begin{figure}[htb!]
	\centering
	%\begin{overpic}
    % \put(30,10){(a)}
    % \end{overpic}
	\includegraphics[width=1.1\textwidth]{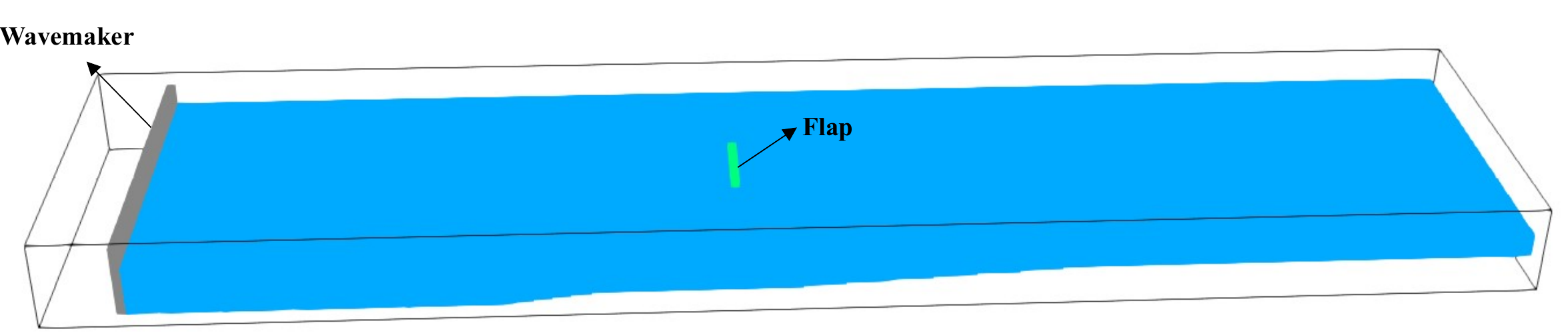} \\
	%\begin{overpic}
    % \put(30,550){(b)}
    % \end{overpic}
	\includegraphics[width=1.1\textwidth]{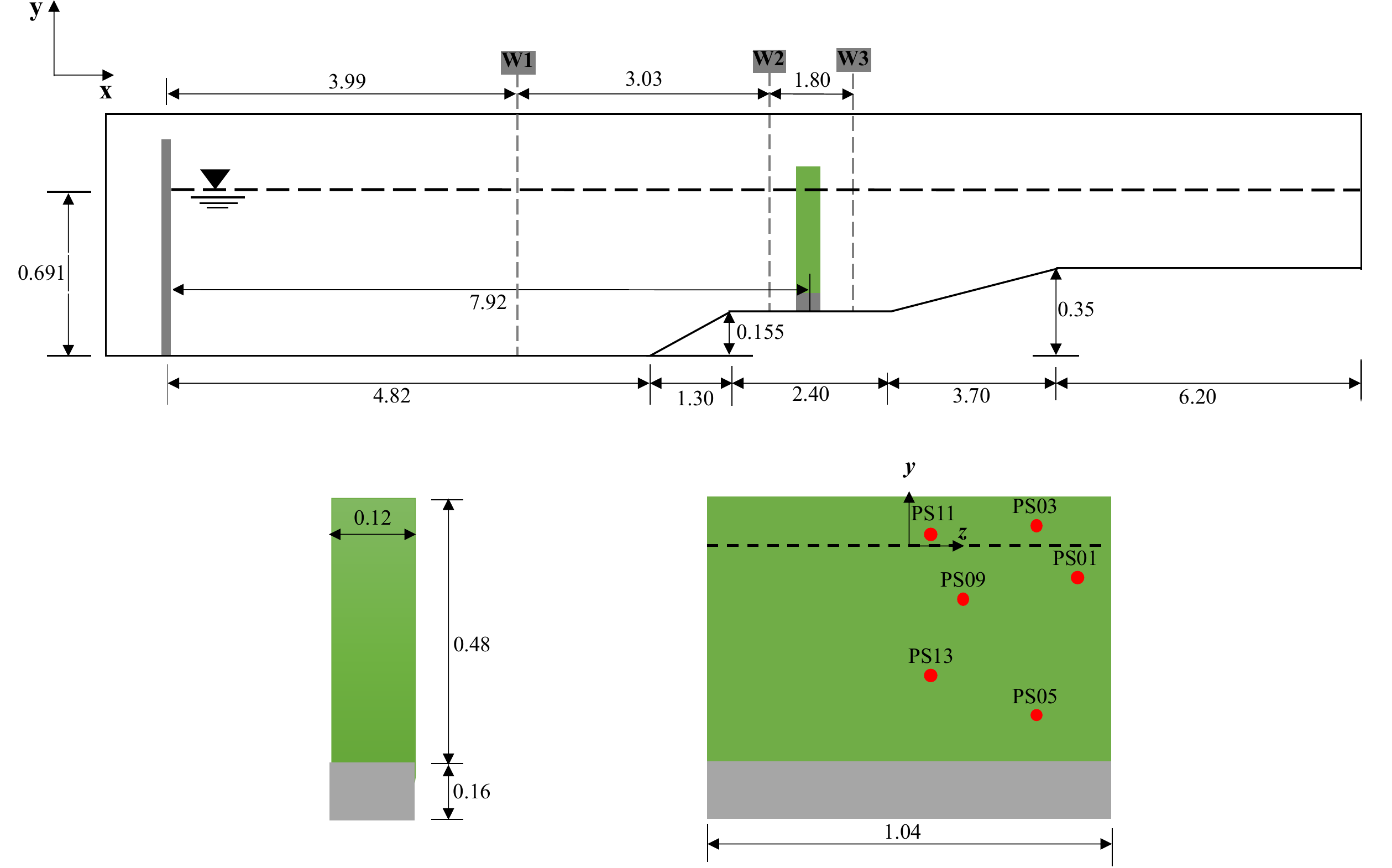} 
	\caption{OWSC: The initial sketch.}
	\label{figs:owscsketch}
\end{figure}

In this section, we apply the present WKGC scheme to study wave interaction with an oscillating wave surge converter (OWSC) which has been recognized as a promising wave energy converter (WEC) and numerically and experimentally studied in the literature \cite{53,54,64}. In this work, both 2D and 3D simulations are conducted. The schematic is shown in Figure.\ref{figs:owscsketch} where the wave tank dimension is 18.2 (length)$\times$ 4.58 (width) $\times$ 1.0 (height) $m^3$ with 1:25 scale \cite{64}. The OWSC device is simplified as a flap with a height of 0.48 m, a width of 1.04 m and a thickness of 0.12 m, hinged to a 0.16 m high base. The mass of the flap is 33 kg and its angular inertia is $1.84~kg/ m^2$. For the coupling of the fluid solver and the rigid-body dynamics, we refer to Ref.\cite{64} for more details. To discretize the system, the particle resolution is 0.03 m resulting in 13104 particles and 2.19 million particles for the 2D and 3D discretization, respectively. Three wave gauges and six pressure sensors are applied to measure the wave elevation and impact pressure on the flap as shown in Figure.\ref{figs:owscsketch}. The positions of pressure sensors are presented in Table \ref{table:PressureSensorPosition}. 

\begin{table*}[htb!]
	\centering
	\caption{OWSC: Positions of the pressure sensors on the front flap face. The position along the z-axis is measured from the center of the device, and $y = 0$ denotes the mean water level.}
	\begin{tabular}{cccccc}
		\hline
		No.    & $y$-axis (m)  & $z$-axis (m)     &  No.    & $y$-axis (m)  & $z$-axis (m)  \\ 
		\hline
		PS01  &$-0.046$       & $0.468$   & PS09  &$-0.117$       & $0.156$ \\ 
		%\hline
		PS03  &$ 0.050$       & $0.364$   & PS11  &$ 0.025$       & $0.052$ \\ 
		%\hline
		PS05  &$-0.300$       & $0.364$   & PS13  &$-0.239$       & $0.052$ \\ 
		\hline
	\end{tabular}
\label{table:PressureSensorPosition}
\end{table*}

For the wave making, we apply the piston-type wave maker to generate the regular wave where the displacement of the wave maker is obtained from linear wavemaker theory \cite{55}
\begin{equation}
\mathbf{r}=\frac{1}{2}Ssin\left(ft+\phi\right),\label{eq24}\\
\end{equation}
where $S$ is the wave stroke, $f$ the wave frequency and $\phi$ the initial phase. Also the wave stroke is given by
\begin{equation}
S=\frac{0.5Hkg/\omega^2}{sinh\left(kh_0\right)cosh\left(kh_0\right)}\left(sinh\left(2kh_0\right)+2kh_0\right),\label{eq25}\\
\end{equation}
where $H$ is the wave height, $k$ the wave number ($k= 2\pi/\lambda$) and $h_0$ the water depth. To avoid the effect of wave reflection, a damping zone \cite{65} is set at the end of the tank where the particle velocity decays as
\begin{equation}
\mathbf{v}=\mathbf{v}_0\left(1.0-\alpha\Delta t\left(\frac{\mathbf{r}-\mathbf{r}_{a0}}{\mathbf{r}_{ae}-\mathbf{r}_{a0}}\right)\right),\label{eq26}\\
\end{equation}
where $\mathbf{v}_0$ is the velocity of fluid particles at the entry of the sponge layer, $\alpha = 5.0$ the damping coefficient, $\mathbf{r}_{a0}$ and $\mathbf{r}_{ae}$ are the beginning and end position of the sponge layer, respectively. 

\begin{figure}[htb!]
	\centering
	\includegraphics[width=0.8\textwidth]{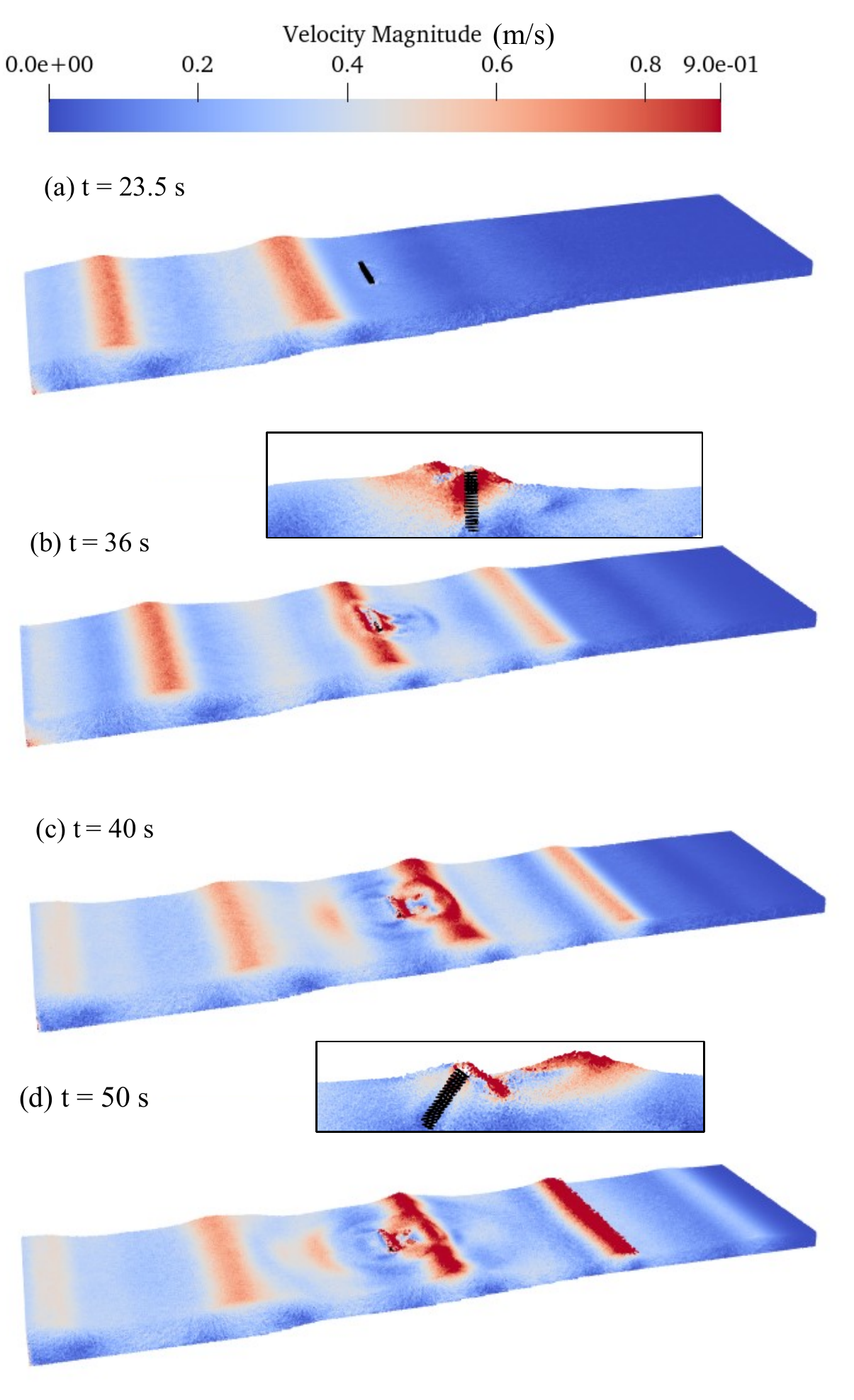} \\
	\caption{OWSC: Snapshots of the free-surface profile colored by velocity magnitude.}
	\label{figs:OWSCVelocityContour}
\end{figure}

Figure.\ref{figs:OWSCVelocityContour} presents snapshots of the free-surface profile colored by velocity magnitude. It can be noted that the present method can capture the violent free-surface elevation involving impacting break and re-entry, and produce a smooth velocity field. The cutting slice along the center of the wave tank portrays the wave reflection and breaking around the flap. 

\begin{figure}[htb!]
	\centering
	\includegraphics[width=1.0\textwidth]{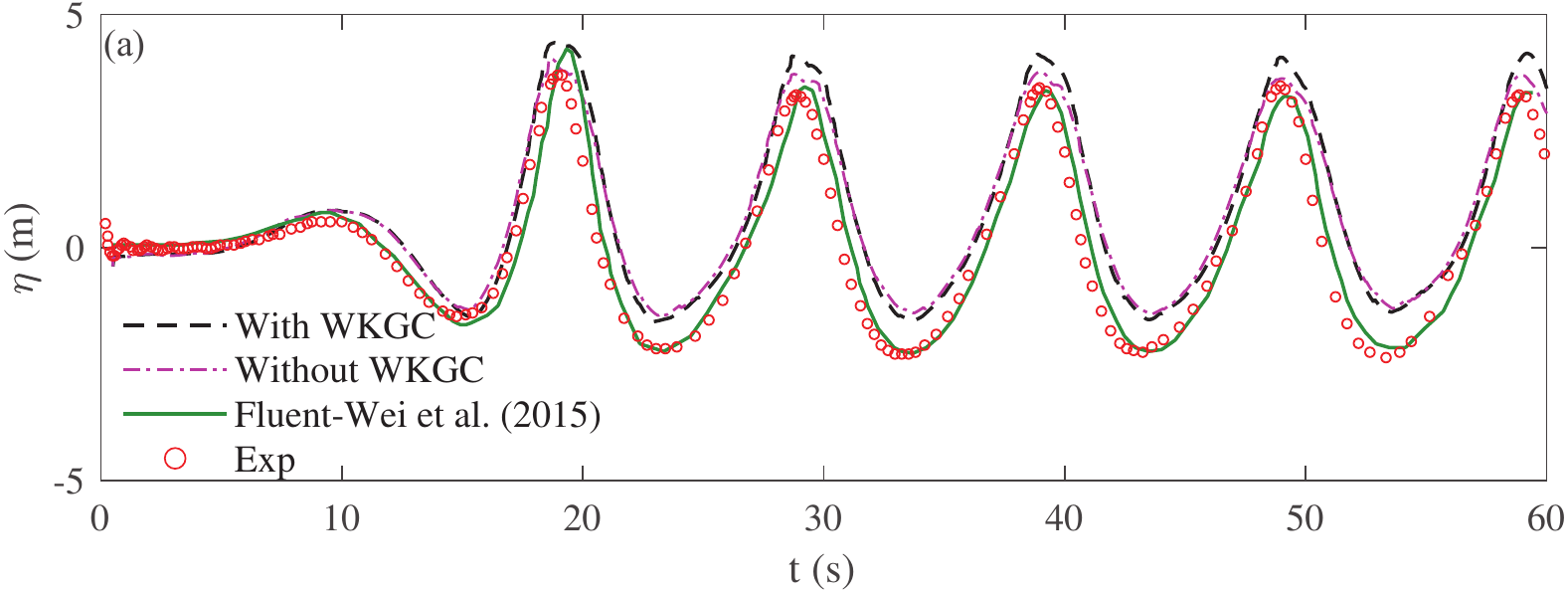} \\
	\includegraphics[width=1.0\textwidth]{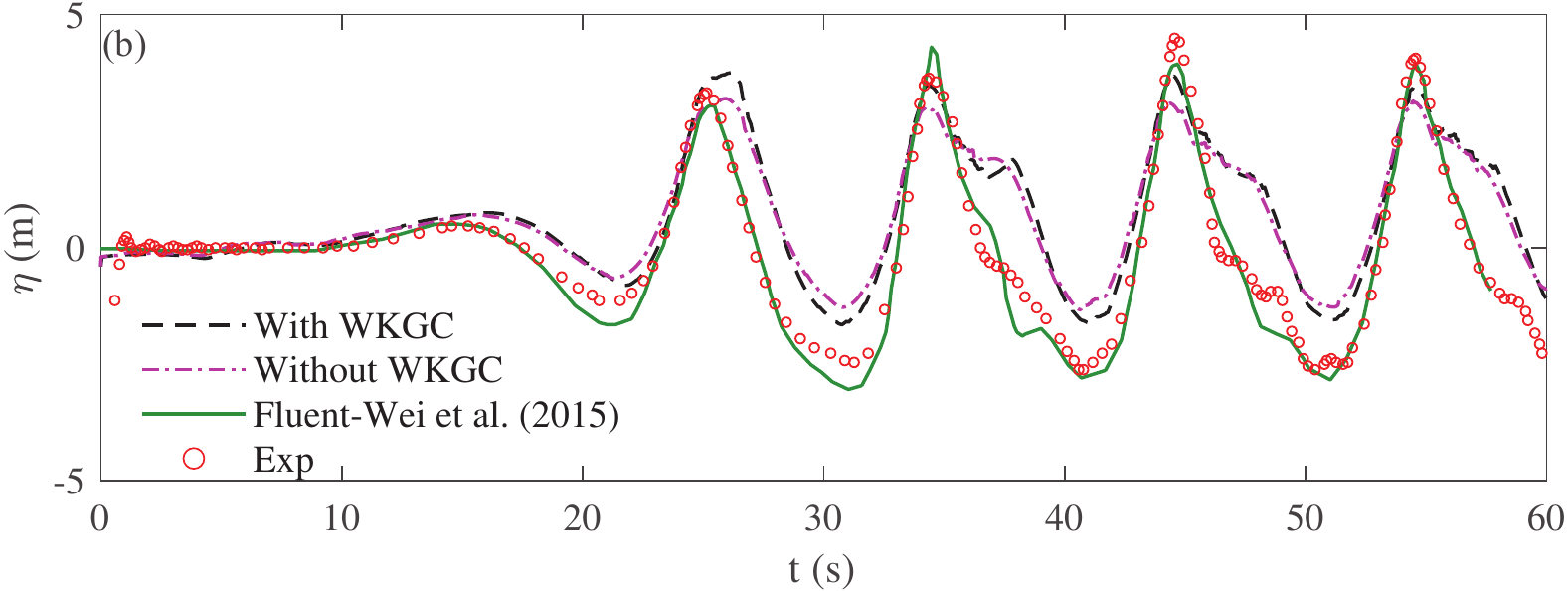} \\
	\includegraphics[width=1.0\textwidth]{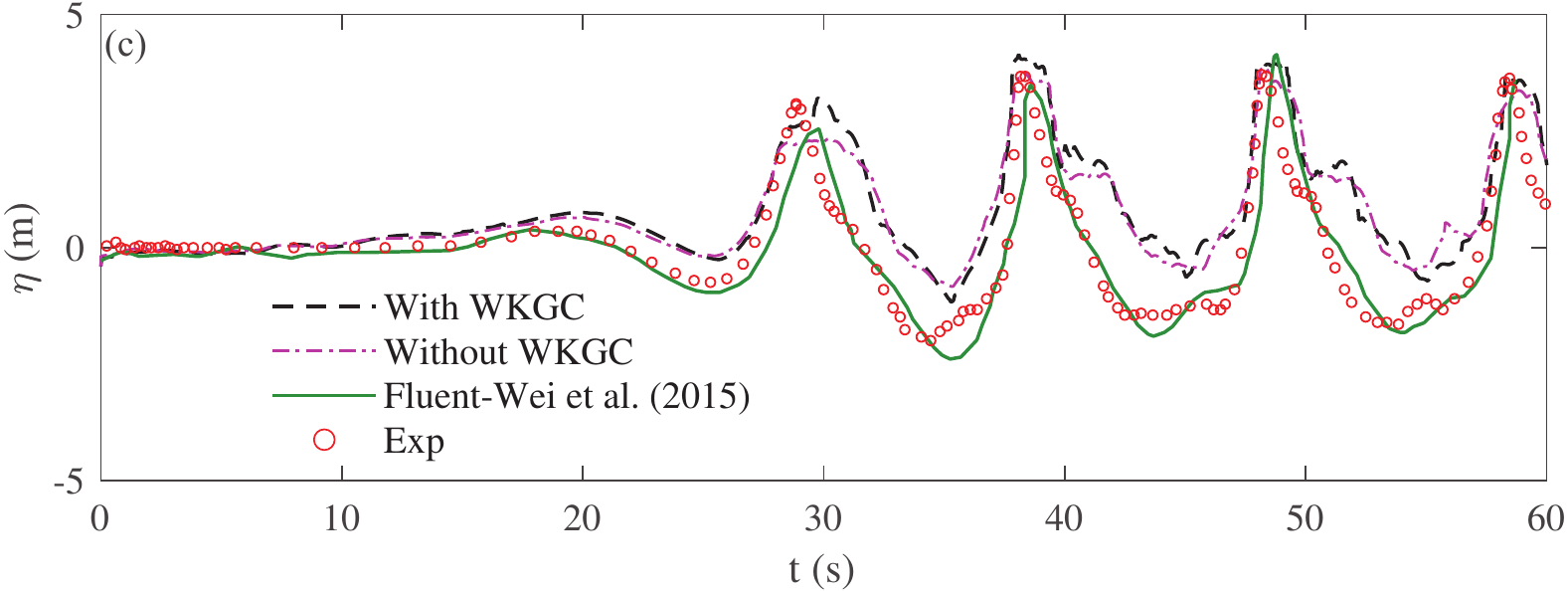} \\
	\caption{OWSC: Time history of the free-surface elevation at W1(a), W2(b) and W3(c).}
	\label{figs:OWSC-Elevation}
\end{figure}

Figure.\ref{figs:OWSC-Elevation} shows the time history of the wave elevation. The present results exhibit an improvement compared to the results obtained without the KGC scheme and demonstrate reasonable agreement with the experimental data. While some discrepancies in wave crest and phase are noted. One possible reason is that wave reflection and breaking occur around the flap, as also shown in Figure.\ref{figs:OWSCVelocityContour}. Another reason could be attributed to the absence of a turbulence model, which may strongly affect the interaction between wave and flap. The overestimate of wave crests is due to the less dissipation after utilizing the WKGC scheme. While Wei et al.\cite{54} adopted the standard k-$\varepsilon$ turbulence model, which introduces more dissipation to the simulation.

%Another reason is the lack of a turbulence model, which may strongly affect the interaction between wave and flap. Note that the standard k-$\varepsilon$ turbulence model is adopted in Wei et al.\cite{54}.

\begin{figure}[htb!]
	\centering
	\includegraphics[width=1.0\textwidth]{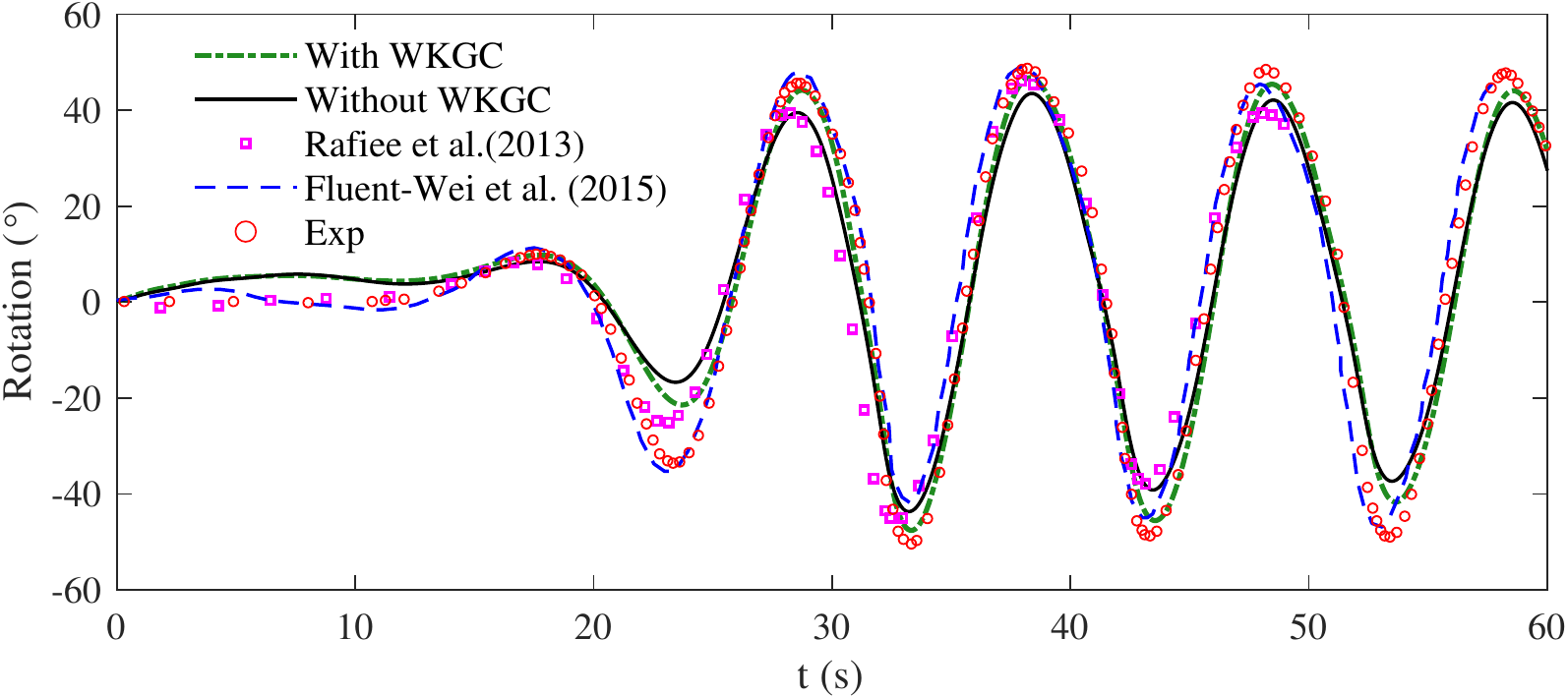} \\
	\caption{OWSC: Time evolution of the flap rotation.}
	\label{figs:OWSC-Rotation}
\end{figure}

Figure.\ref{figs:OWSC-Rotation} presents the time history of the rotation of the flap obtained by the present method and the method without WKGC, and its comparison with other published numerical results \cite{53,54} and experimental data \cite{54}. The present numerical results show clear improvements and agree better with the experimental data compared with that without WKGC. Compared with the results obtained by Wei et al.\cite{54}, the SPH predictions generally underestimated rotation due to the lack of a turbulence model. Compared with SPH results in Refs. \cite{53,64}, the WKGC scheme can predict flap motion more accurately with less deviation from the experimental rotation crest.

\begin{landscape}

\begin{figure}[htb!]
	\centering
        \includegraphics[width=0.65\textwidth]{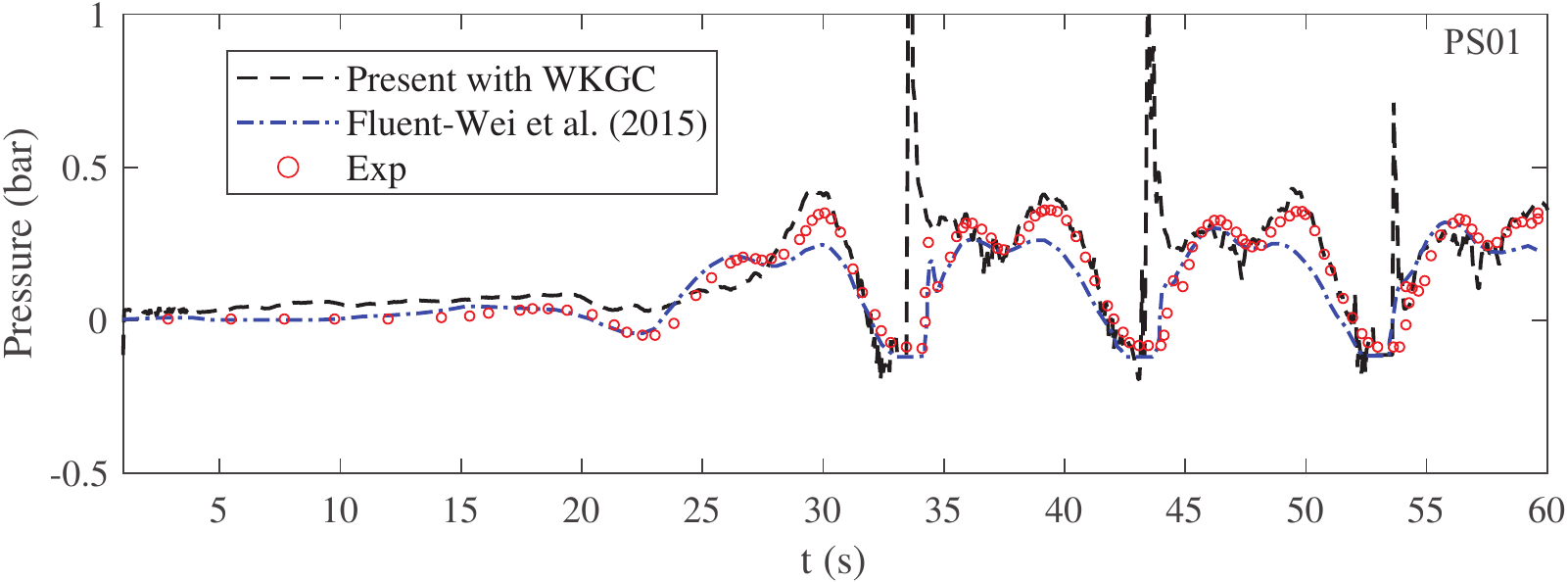}\vspace{1pt}
        \includegraphics[width=0.65\textwidth]{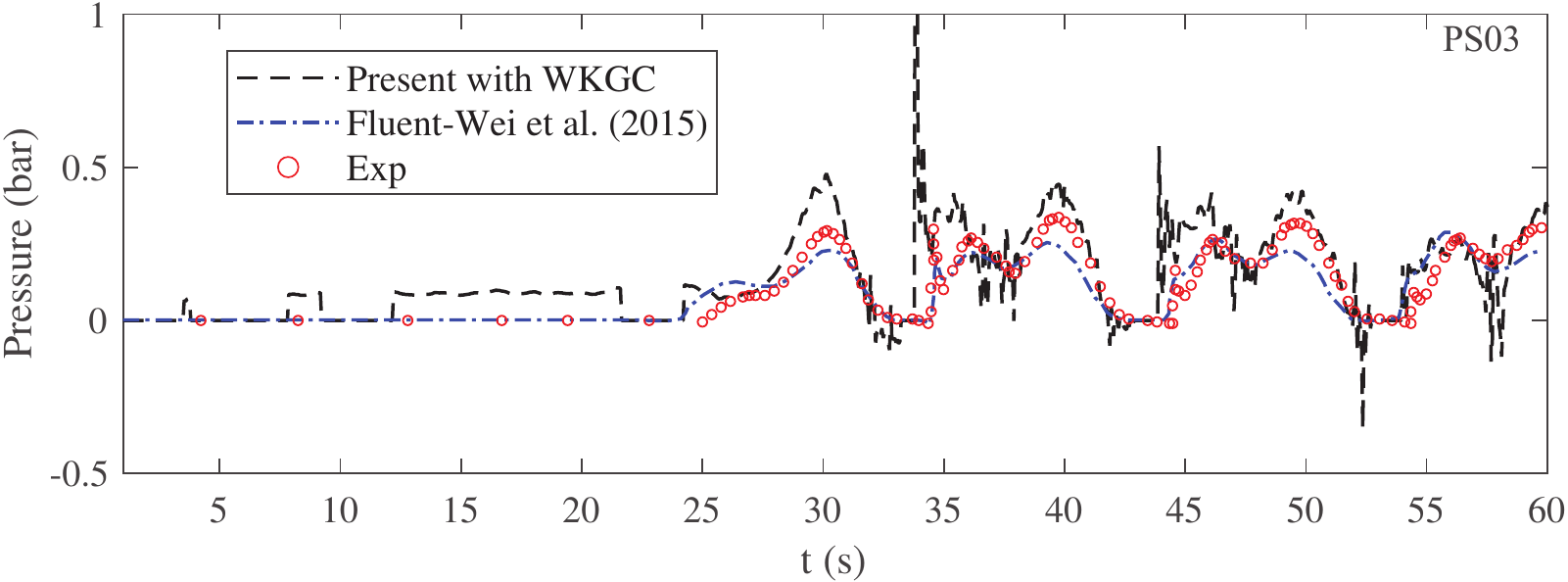}\\
        \includegraphics[width=0.65\textwidth]{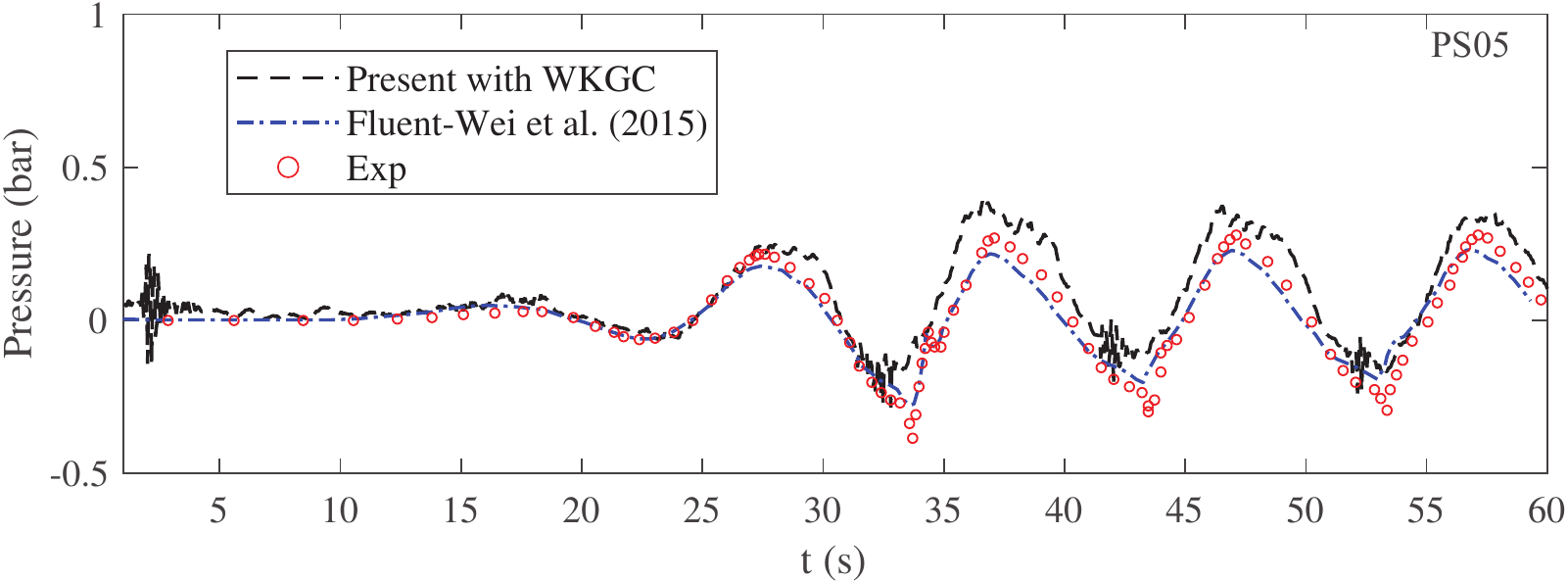}\vspace{1pt}        
        \includegraphics[width=0.65\textwidth]{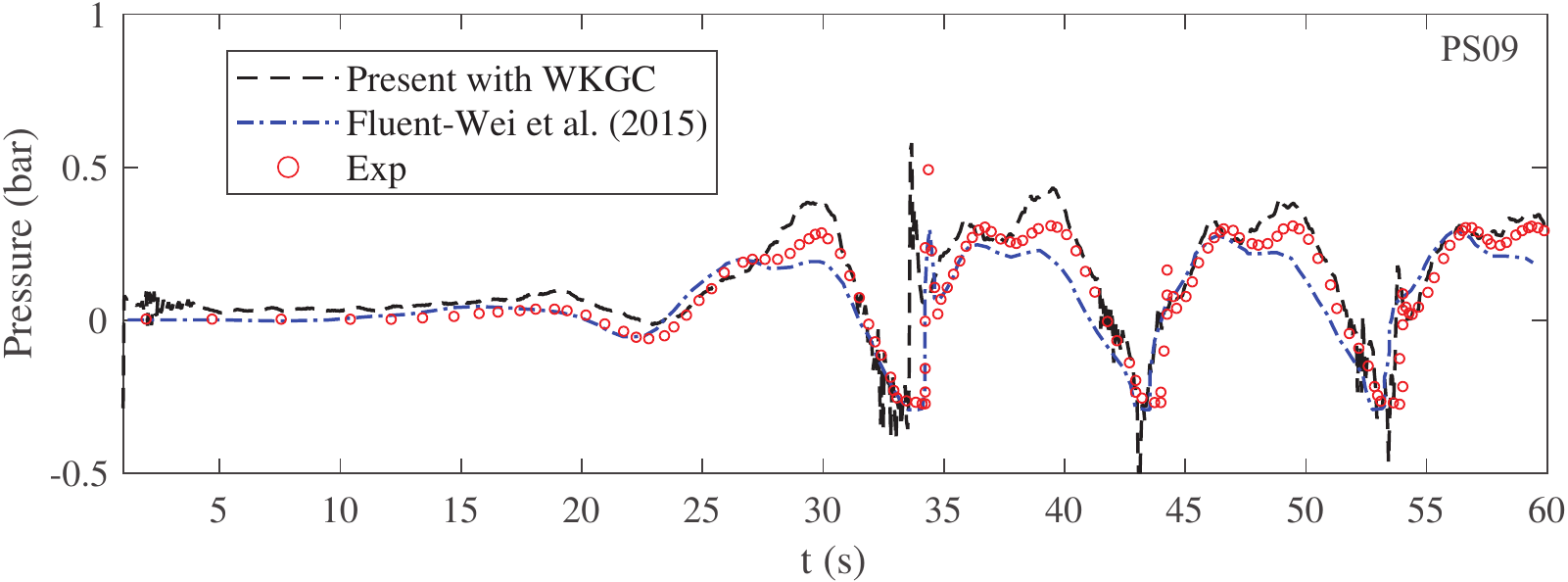}\\
        \includegraphics[width=0.65\textwidth]{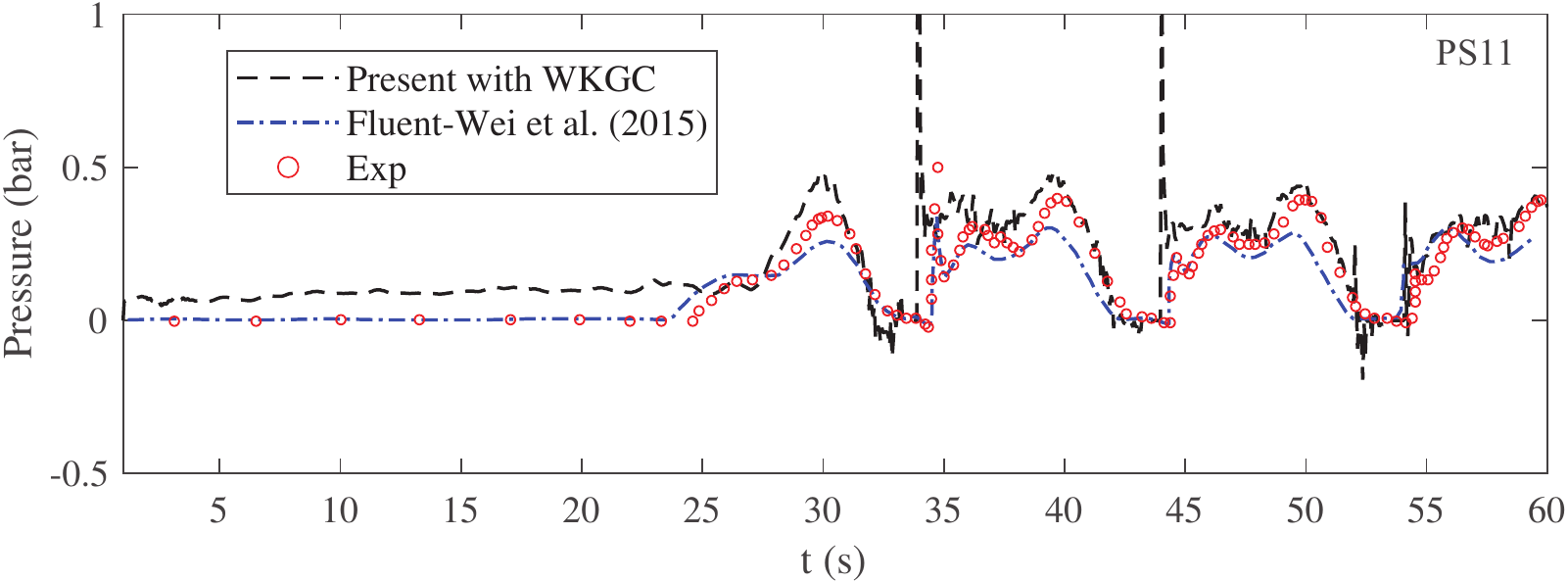}\vspace{1pt}
        \includegraphics[width=0.65\textwidth]{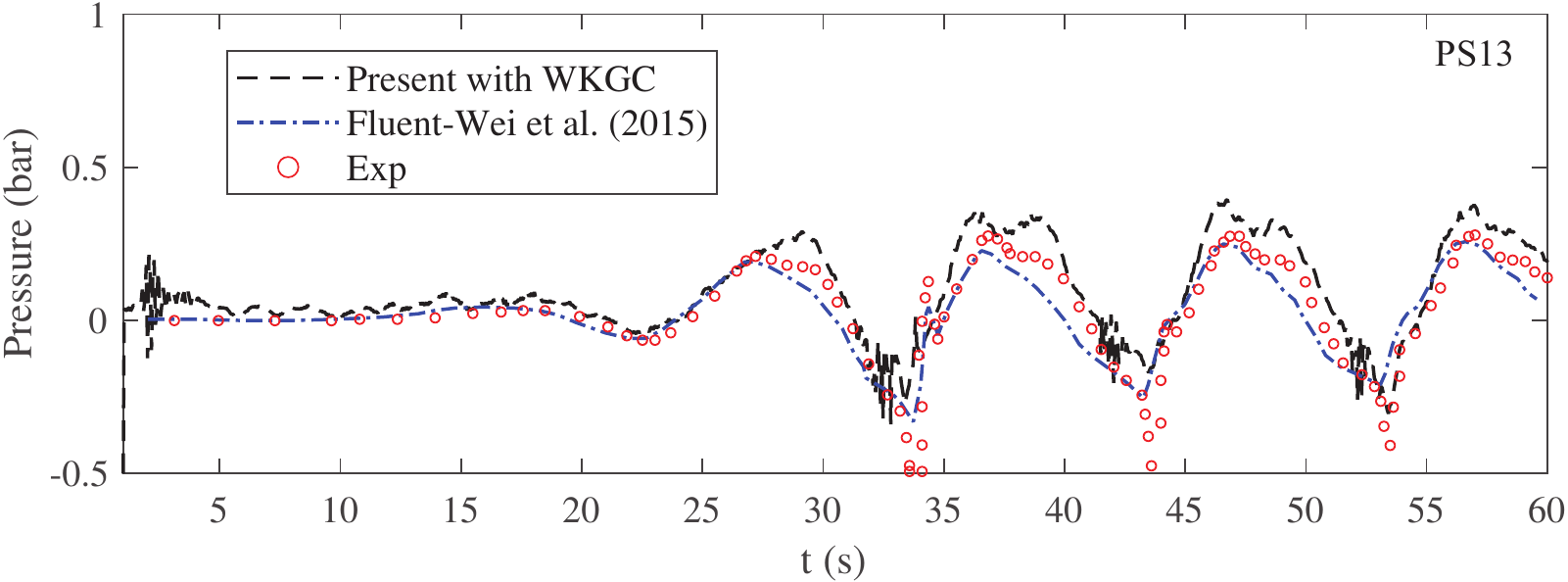}\\
       \caption{OWSC: Time history of pressure on the flap obtained by the present method and its comparison with that of Wei et al.\cite{54}.}
	\label{figs:OWSCPressure3D}
\end{figure}
\end{landscape}

Figure.\ref{figs:OWSCPressure3D} presents the time history of the probed pressure on the flap obtained by the present method and its comparison with that of Wei et al. \cite{54} and experimental data \cite{54}. Generally, the results show good agreement. For pressure sensors PS01, PS03, PS09 and PS11, the slamming pressure can be well captured except for some high-frequency pressure oscillations. Compared to Wei's numerical results, the present method effectively captures the double peak and agrees better with experimental data. Similar to the numerical results from Wei et al. \cite{54}, the pressure drops of the present method are underestimated compared with the experimental observation for sensors PS05 and PS13. This difference originates mainly from the weakly compressible assumption in the present method and the splash passing the flap usually accompanies the air entrainment which is not considered in the numerical model.

The computational efficiency of the present method with and without WKGC is analyzed in Table \ref{table: CPUTime}. The 3D simulations are carried out on an Intel(R) Xeon(R) Platinum 9242 CPU @ 2.30GHz with 48 cores and the 2D ones are on an Intel Core i7-9750H laptop with 6 cores. For the 3D simulations, the introduction of the present WKGC slightly increases the computational wall clock time while the induced extra computational efforts are negligible for 2D simulations.
%%%%%%%%%%
% Section
%%%%%%%%%%%%%%%%%%%%%%%%%%%%%%%%%%%%%%%%%%%%%%%%%%%%%%%%%%%%%
\section{Conclusions}
\label{Conclusions}

In this paper, we proposed an efficient, robust and simple WKGC scheme to address the issue of numerical instability and computational efficiency for introducing the KGC scheme in the Riemann-SPH method. The underlying principle is to introduce a weighted value of the original KGC matrix and the identity one, implement it in a particle-average manner and cooperate it into a dual-criteria time stepping framework. Extensive examples, including standing wave, oscillating drop, dam-break flow and wave interacting with an OWSC, are investigated to demonstrate that the present WKGC scheme can stably resolve violent free surface flows, reproduce a smooth pressure field, reduce numerical dissipation meanwhile induces limited extra computational efforts. 
%%%%%%%%%%%%%%%%%%%%%%%%%%%%%%%%%%%%%%%%%%%

\begin{table}[t]
  \centering
  \caption{OWSC: The CPU wall-clock time for both 2D and 3D simulations.}
  \resizebox{\textwidth}{!}{
  \begin{tabularx}{1.2\textwidth}{p{7em}<{\centering}p{6em}<{\centering}p{6em}<{\centering}p{6em}<{\centering}p{9em}<{\centering}}
    \hline
    Model & Computational time & Physical time & Particle number& Device Info\\
    \hline
   With WKGC (3D) & 4.58 h & 13.8 s & 2.19 million & 48 cores Intel Xeon Platinum 9242\\
   Without WKGC (3D) & 3.97 h & 13.8 s & 2.19 million & 48 cores Intel Xeon Platinum 9242\\
   With WKGC (2D) & 165.06 s & 13.8 s & 13104 & 6 cores Intel Core i7-9750H\\
   Without WKGC (2D)& 153.78 s & 13.8 s & 13104 & 6 cores Intel Core i7-9750H\\
   \hline
\end{tabularx}
}
\label{table: CPUTime}
\end{table}
  
  %%%%%%%%%%%%%%%%%%%%%%%%%%%%%%%%%%%%%%%%%%%%%%%%%%%%%%%%%%%%%
  %
  % Section
  %
  %%%%%%%%%%%%%%%%%%%%%%%%%%%%%%%%%%%%%%%%%%%%%%%%%%%%%%%%%%%%%
  \section*{CRediT authorship contribution statement}
   {\bfseries  Yaru Ren:} Investigation, Methodology, Visualization, Formal analysis, Writing - original draft, Writing - review \& editing; 
   {\bfseries  Pengzhi Lin:} Investigation, Supervision, Writing - review \& editing;
  {\bfseries  Chi Zhang:} Investigation, Methodology, Formal analysis, Writing - review \& editing; 
  {\bfseries  Xiangyu Hu:} Investigation, Supervision, Writing - review \& editing.
  %%%%%%%%%%%%%%%%%%%%%%%%%%%%%%%%%%%%%%%%%%%%%%%%%%%%%%%%%%%%%
  %
  % Section
  %
  %%%%%%%%%%%%%%%%%%%%%%%%%%%%%%%%%%%%%%%%%%%%%%%%%%%%%%%%%%%%%
  \section*{Declaration of competing interest }
  The authors declare that they have no known competing financial interests 
  or personal relationships that could have appeared to influence the work reported in this paper.
  %%%%%%%%%%%%%%%%%%%%%%%%%%%%%%%%%%%%%%%%%%%%%%%%%%%%%%%%%%%%%
  %
  % Section
  %
  %%%%%%%%%%%%%%%%%%%%%%%%%%%%%%%%%%%%%%%%%%%%%%%%%%%%%%%%%%%%%
  \section{Acknowledgement}
  C. Zhang and X.Y. Hu would like to express their gratitude to Deutsche Forschungsgemeinschaft (DFG) 
  for their sponsorship of this research under grant numbers  HU1527/12-4. 
 
%%%%%%%%%%%%%%%%%%%%%%%%%%%%%%%%%%%%%%%%%%%%%%%%%%%%%%%%%%%%%
%
% Section
%
%%%%%%%%%%%%%%%%%%%%%%%%%%%%%%%%%%%%%%%%%%%%%%%%%%%%%%%%%%%%%
%\section*{Appendix}
%\label{appendix}
%%%%%%%%%%%%%%%%%%%%%%%%%%%%%%%%%%%%%%%%%%%%%%%%%%%%%%%%%%%%%
% Section
%%%%%%%%%%%%%%%%%%%%%%%%%%%%%%%%%%%%%%%%%%%%%%%%%%%%%%%%%%%%%
%\subsection*{Appendix A : The pairwise splitting approach}\label{appendix A}

%\section*{References}
\bibliographystyle{elsarticle-num}
\bibliography{correction-matrix}

\begin{thebibliography}{10}
\expandafter\ifx\csname url\endcsname\relax
  \def\url#1{\texttt{#1}}\fi
\expandafter\ifx\csname urlprefix\endcsname\relax\def\urlprefix{URL }\fi
\expandafter\ifx\csname href\endcsname\relax
  \def\href#1#2{#2} \def\path#1{#1}\fi

\bibitem{22}
C.~Zhang, M.~Rezavand, X.~Hu, Dual-criteria time stepping for weakly
  compressible smoothed particle hydrodynamics, Journal of Computational
  Physics 404 (2020) 109135.

\bibitem{1}
H.~Gotoh, A.~Khayyer, On the state-of-the-art of particle methods for coastal
  and ocean engineering, Coastal Engineering Journal 60~(1) (2018) 79--103.

\bibitem{2}
M.~Luo, A.~Khayyer, P.~Lin, Particle methods in ocean and coastal engineering,
  Applied Ocean Research 114 (2021) 102734.

\bibitem{3}
C.~Zhang, Y.-j. Zhu, D.~Wu, N.~A. Adams, X.~Hu, Smoothed particle
  hydrodynamics: Methodology development and recent achievement, Journal of
  Hydrodynamics 34~(5) (2022) 767--805.

\bibitem{67}
D.~Violeau, B.~D. Rogers, Smoothed particle hydrodynamics (sph) for
  free-surface flows: past, present and future, Journal of Hydraulic Research
  54~(1) (2016) 1--26.

\bibitem{68}
C.~Altomare, A.~J. Crespo, J.~M. Dom{\'\i}nguez, M.~G{\'o}mez-Gesteira,
  T.~Suzuki, T.~Verwaest, Applicability of smoothed particle hydrodynamics for
  estimation of sea wave impact on coastal structures, Coastal Engineering 96
  (2015) 1--12.

\bibitem{57}
V.~Springel, Smoothed particle hydrodynamics in astrophysics, Annual Review of
  Astronomy and Astrophysics 48 (2010) 391--430.

\bibitem{56}
J.~Wang, D.~Chan, Frictional contact algorithms in sph for the simulation of
  soil--structure interaction, International Journal for Numerical and
  Analytical Methods in Geomechanics 38~(7) (2014) 747--770.

\bibitem{4}
P.~Guilcher, G.~Ducorzet, B.~Alessandrini, P.~Ferrant, Water wave propagation
  using sph models, in: Proceedings 2 nd International Spheric Workshop, 2007,
  pp. 119--122.

\bibitem{27}
T.~Kanehira, M.~L. McAllister, S.~Draycott, T.~Nakashima, D.~M. Ingram, T.~S.
  van~den Bremer, H.~Mutsuda, The effects of smoothing length on the onset of
  wave breaking in smoothed particle hydrodynamics (sph) simulations of highly
  directionally spread waves, Computational Particle Mechanics (2022) 1--17.

\bibitem{6}
A.~Khayyer, H.~Gotoh, Y.~Shimizu, K.~Gotoh, On enhancement of energy
  conservation properties of projection-based particle methods, European
  Journal of Mechanics-B/Fluids 66 (2017) 20--37.

\bibitem{5}
V.~Zago, L.~J. Schulze, G.~Bilotta, N.~Almashan, R.~Dalrymple, Overcoming
  excessive numerical dissipation in sph modeling of water waves, Coastal
  Engineering 170 (2021) 104018.

\bibitem{50}
S.~Marrone, M.~Antuono, A.~Colagrossi, G.~Colicchio, D.~Le~Touz{\'e},
  G.~Graziani, $\delta$-sph model for simulating violent impact flows, Computer
  Methods in Applied Mechanics and Engineering 200~(13-16) (2011) 1526--1542.

\bibitem{28}
P.~Sun, A.~Colagrossi, S.~Marrone, A.~Zhang, The $\delta$plus-sph model: Simple
  procedures for a further improvement of the sph scheme, Computer Methods in
  Applied Mechanics and Engineering 315 (2017) 25--49.

\bibitem{21}
C.~Zhang, X.~Hu, N.~A. Adams, A weakly compressible sph method based on a
  low-dissipation riemann solver, Journal of Computational Physics 335 (2017)
  605--620.

\bibitem{30}
Y.~Ren, P.~Lin, A.~Khayyer, M.~Luo, Comparative analysis of three smoothed
  particle hydrodynamics methods in modeling free-surface flows, International
  Journal of Offshore and Polar Engineering 32~(03) (2022) 267--274.

\bibitem{7}
A.~Colagrossi, A.~Souto-Iglesias, M.~Antuono, S.~Marrone,
  Smoothed-particle-hydrodynamics modeling of dissipation mechanisms in gravity
  waves, Physical Review E 87~(2) (2013) 023302.

\bibitem{13}
P.~Randles, L.~D. Libersky, Smoothed particle hydrodynamics: some recent
  improvements and applications, Computer methods in applied mechanics and
  engineering 139~(1-4) (1996) 375--408.

\bibitem{14}
J.~Bonet, T.-S. Lok, Variational and momentum preservation aspects of smooth
  particle hydrodynamic formulations, Computer Methods in applied mechanics and
  engineering 180~(1-2) (1999) 97--115.

\bibitem{15}
A.~Khayyer, H.~Gotoh, S.~Shao, Corrected incompressible sph method for accurate
  water-surface tracking in breaking waves, Coastal Engineering 55~(3) (2008)
  236--250.

\bibitem{16}
H.~Wen, B.~Ren, X.~Yu, An improved sph model for turbulent hydrodynamics of a
  2d oscillating water chamber, Ocean Engineering 150 (2018) 152--166.

\bibitem{17}
Y.~Xiao, X.~Hong, Z.~Tang, Normalized sph without boundary deficiency and its
  application to transient solid mechanics problems, Meccanica 55~(11) (2020)
  2263--2283.

\bibitem{66}
J.~P. Vila, Sph renormalized hybrid methods for conservation laws: applications
  to free surface flows, in: Meshfree methods for partial differential
  equations II, 2005, pp. 207--229.

\bibitem{18}
S.~L.~Z. V., B.~G., D.~R.A., Localized kernel gradient correction for sph
  simulations of water wave propagation, Proceedings of the 16th SPHERIC
  International Workshop (2022).

\bibitem{20}
Z.~Zhang, M.~Liu, A decoupled finite particle method for modeling
  incompressible flows with free surfaces, Applied Mathematical Modelling 60
  (2018) 606--633.

\bibitem{43}
J.~J. Monaghan, Smoothed particle hydrodynamics, Annual review of astronomy and
  astrophysics 30 (1992) 543--574.

\bibitem{24}
T.~Ye, D.~Pan, C.~Huang, M.~Liu, Smoothed particle hydrodynamics (sph) for
  complex fluid flows: Recent developments in methodology and applications,
  Physics of Fluids 31~(1) (2019) 011301.

\bibitem{19}
H.~Xiaoting, S.~Pengnan, L.~Hongguan, Z.~Shiyun, Development of a numerical
  wave tank with a corrected smoothed particle hydrodynamics scheme to reduce
  nonphysical energy dissipation, Chinese Journal of Theoretical and Applied
  Mechanics 54~(6) (2022) 1502--1515.

\bibitem{44}
T.~Whittaker, D.~Collier, M.~Folley, M.~Osterried, A.~Henry, M.~Crowley, The
  development of oyster—a shallow water surging wave energy converter, in:
  Proceedings of the 7th European wave and tidal energy conference, 2007, pp.
  11--14.

\bibitem{26}
G.~Wu, R.~E. Taylor, Finite element analysis of two-dimensional non-linear
  transient water waves, Applied Ocean Research 16~(6) (1994) 363--372.

\bibitem{25}
A.~Khayyer, Y.~Shimizu, T.~Gotoh, H.~Gotoh, Enhanced resolution of the
  continuity equation in explicit weakly compressible sph simulations of
  incompressible free-surface fluid flows, Applied Mathematical Modelling
  (2022).

\bibitem{39}
M.~Antuono, S.~Marrone, A.~Colagrossi, B.~Bouscasse, Energy balance in the
  $\delta$-sph scheme, Computer Methods in Applied Mechanics and Engineering
  289 (2015) 209--226.

\bibitem{40}
I.~Hammani, S.~Marrone, A.~Colagrossi, G.~Oger, D.~Le~Touz{\'e}, Detailed study
  on the extension of the $\delta$-sph model to multi-phase flow, Computer
  Methods in Applied Mechanics and Engineering 368 (2020) 113189.

\bibitem{41}
J.~J. Monaghan, A.~Rafiee, A simple sph algorithm for multi-fluid flow with
  high density ratios, International Journal for Numerical Methods in Fluids
  71~(5) (2013) 537--561.

\bibitem{38}
H.~Xiaoting, S.~Pengnan, L.~Hongguan, Z.~Shiyun, Development of a numerical
  wave tank with a corrected smoothed particle hydrodynamics scheme to reduce
  nonphysical energy dissipation, Chinese Journal of Theoretical and Applied
  Mechanics 54~(6) (2022) 1502--1515.

\bibitem{59}
A.~Colagrossi, B.~Bouscasse, M.~Antuono, S.~Marrone, Particle packing algorithm
  for sph schemes, Computer Physics Communications 183~(8) (2012) 1641--1653.

\bibitem{48}
M.~Jandaghian, H.~M. Siaben, A.~Shakibaeinia, Stability and accuracy of the
  weakly compressible sph with particle regularization techniques, European
  Journal of Mechanics-B/Fluids 94 (2022) 314--333.

\bibitem{49}
A.~Krimi, M.~Jandaghian, A.~Shakibaeinia, A wcsph particle shifting strategy
  for simulating violent free surface flows, Water 12~(11) (2020) 3189.

\bibitem{45}
B.~Buchner, Green water on ship-type offshore structures, Ph.D. thesis, Delft
  University of Technology Delft, The Netherlands (2002).

\bibitem{46}
L.~Lobovsk{\`y}, E.~Botia-Vera, F.~Castellana, J.~Mas-Soler, A.~Souto-Iglesias,
  Experimental investigation of dynamic pressure loads during dam break,
  Journal of Fluids and Structures 48 (2014) 407--434.

\bibitem{47}
J.~Martin, W.~Moyce, J.~Martin, W.~Moyce, W.~G. Penney, A.~Price, C.~Thornhill,
  Part v. an experimental study of the collapse of fluid columns on a rigid
  horizontal plane, Philosophical Transactions of the Royal Society of London.
  Series A, Mathematical and Physical Sciences 244~(882) (1952) 325--334.

\bibitem{51}
R.~A, Die fortpflanzung de wasserwellen, Zeitschrift Verein Deutscher
  Ingenieure 36~(33) (1892) 947–954.

\bibitem{53}
A.~Rafiee, B.~Elsaesser, F.~Dias, Numerical simulation of wave interaction with
  an oscillating wave surge converter, in: International Conference on Offshore
  Mechanics and Arctic Engineering, Vol. 55393, American Society of Mechanical
  Engineers, 2013, p. V005T06A013.

\bibitem{54}
Y.~Wei, A.~Rafiee, A.~Henry, F.~Dias, Wave interaction with an oscillating wave
  surge converter, part i: Viscous effects, Ocean Engineering 104 (2015)
  185--203.

\bibitem{64}
C.~Zhang, Y.~Wei, F.~Dias, X.~Hu, An efficient fully lagrangian solver for
  modeling wave interaction with oscillating wave surge converter, Ocean
  Engineering 236 (2021) 109540.

\bibitem{55}
R.~G. Dean, R.~A. Dalrymple, Water wave mechanics for engineers and scientists,
  Vol.~2, world scientific publishing company, 1991.

\bibitem{65}
S.~J. Lind, R.~Xu, P.~K. Stansby, B.~D. Rogers, Incompressible smoothed
  particle hydrodynamics for free-surface flows: A generalised diffusion-based
  algorithm for stability and validations for impulsive flows and propagating
  waves, Journal of Computational Physics 231~(4) (2012) 1499--1523.

\end{thebibliography}
%%%%%%%%%%%%%%%%%%%%%%%%%%%%%%%%%%%%%%%%%%%%%%%%%%%%%%%%%%%%%
%
% Section
%
%%%%%%%%%%%%%%%%%%%%%%%%%%%%%%%%%%%%%%%%%%%%%%%%%%%%%%%%%%%%%
\end{document}